\documentclass[fleqn,usenatbib]{mnras}

\usepackage{newtxtext,newtxmath}

\usepackage[T1]{fontenc}

\newcommand{\orcit}[1]{\protect\href{https://orcid.org/#1}{\protect\includegraphics[width=8pt]{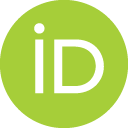}}}

\DeclareRobustCommand{\VAN}[3]{#2}
\let\VANthebibliography\thebibliography
\def\thebibliography{\DeclareRobustCommand{\VAN}[3]{##3}\VANthebibliography}

\usepackage{graphicx}	
\usepackage{amsmath}	

\title[Extreme variability in S4 0954+65]{Extreme photometric and polarimetric variability of blazar S4 0954+65 at its maximum optical and $\gamma$-ray brightness levels}

\author[C. M. Raiteri et al.] { 
C.~M.~Raiteri\orcit{0000-0003-1784-2784}               $^{ 1}$\thanks{E-mail:claudia.raiteri@inaf.it}, 
M.~Villata\orcit{0000-0003-1743-6946}                  $^{ 1}$,
M.~I.~Carnerero\orcit{0000-0001-5843-5515}             $^{ 1}$,
S.~S.~Savchenko\orcit{0000-0003-4147-3851}             $^{ 2,
 3,
 4}$,
S.~O.~Kurtanidze                                       $^{ 5}$,
\newauthor
V.~V.~Vlasyuk                                          $^{ 3}$,
A.~Marchini\orcit{0000-0003-3779-6762}                 $^{ 6}$,
K.~Matsumoto                                           $^{ 7}$,
C.~Lorey                                               $^{ 8}$,
M.~D.~Joner\orcit{0000-0003-0634-8449}                 $^{ 9}$,
K.~Gazeas\orcit{0000-0002-8855-3923}                   $^{10}$,
\newauthor
D.~Carosati                                            $^{11,
12}$,
D.~O.~Mirzaqulov\orcit{0000-0003-0570-6531}            $^{13}$,
J.~A.~Acosta Pulido                                    $^{14,
15}$,
I.~Agudo                                               $^{16}$,
R.~Bachev                                              $^{17}$,
\newauthor
E.~Ben\'itez\orcit{0000-0003-1018-2613}                $^{18}$,
G.~A.~Borman\orcit{0000-0002-7262-6710}                $^{19}$,
P.~Calcidese                                           $^{20}$,
W.~P.~Chen                                             $^{21}$,
G.~Damljanovic\orcit{0000-0002-6710-6868}              $^{22}$,
\newauthor
S.~A.~Ehgamberdiev\orcit{0000-0001-9730-3769}          $^{13,
23}$,
D.~Els\"asser\orcit{0000-0001-6796-3205}               $^{24,
 8}$,
M.~Feige                                               $^{ 8}$,
A.~Frasca\orcit{0000-0002-0474-0896}                   $^{25}$,
H.~Gaur                                                $^{26}$,
T.~S.~Grishina\orcit{0000-0002-3953-6676}              $^{ 2}$,
\newauthor
A.~C.~Gupta\orcit{0000-0002-9331-4388}                 $^{26}$,
D.~Hiriart\orcit{0000-0002-4711-7658}                  $^{27}$,
M.~Holland                                             $^{ 9}$,
B.~Horst                                               $^{ 8}$,
S.~Ibryamov                                            $^{28}$,
R.~Z.~Ivanidze                                         $^{ 5}$,
J.~Jensen                                              $^{ 9}$,
\newauthor
V.~Jithesh                                             $^{29}$,
M.~D.~Jovanovic\orcit{0000-0003-4298-3247}             $^{22}$,
S.~Kiehlmann                                           $^{30,
31}$,
G.~N.~Kimeridze                                        $^{ 5}$,
S.~Kishore                                             $^{26}$,
\newauthor
E.~N.~Kopatskaya\orcit{0000-0001-9518-337X}            $^{ 2}$,
O.~M.~Kurtanidze                                       $^{ 5,
32}$,
E.~G.~Larionova\orcit{0000-0002-2471-6500}             $^{ 2}$,
H.~C.~Lin                                              $^{21}$,
K.~Mannheim                                            $^{33,
 8}$,
\newauthor
C.~Marinelli\orcit{0000-0002-3596-4307}                $^{34}$,
J.~Moreira Reyes                                       $^{35}$,
D.~A.~Morozova\orcit{0000-0002-9407-7804}              $^{ 2}$,
M.~G.~Nikolashvili                                     $^{ 5}$,
D.~Reinhart                                            $^{ 8,
33}$,
\newauthor
F.~D.~Romanov\orcit{0000-0002-5268-7735}               $^{36,
37}$,
E.~Semkov                                              $^{17}$,
J.~Seufert                                             $^{ 8}$,
E.~V.~Shishkina                                        $^{ 2}$,
L.~A.~Sigua                                            $^{ 5}$,
R.~Skalidis                                            $^{38}$,
\newauthor
O.~I.~Spiridonova                                      $^{ 3}$,
M.~Stojanovic\orcit{0000-0002-4105-7113}               $^{22}$,
A.~Strigachev                                          $^{17}$,
Y.~V.~Troitskaya\orcit{0000-0002-9907-9876}            $^{ 2}$,
I.~S.~Troitskiy\orcit{0000-0002-4218-0148}             $^{ 2}$,
\newauthor
A.~Tsai                                                $^{21}$,
A.~A.~Vasilyev\orcit{0000-0002-8293-0214}              $^{ 2}$,
O.~Vince                                               $^{22}$,
K.~Vrontaki                                            $^{10}$,
K.~Wani                                                $^{26}$,
D.~Watts                                               $^{ 9}$,
and A.~V.~Zhovtan                                       $^{19}$\\
{\it Affiliations are listed at the end of the paper}        
 }

\date{Accepted XXX. Received YYY; in original form ZZZ}

\pubyear{2015}

\begin{document}
\label{firstpage}
\pagerange{\pageref{firstpage}--\pageref{lastpage}}
\maketitle

\begin{abstract}
In 2022 the BL Lac object S4~0954+65 underwent a major variability phase, reaching its historical maximum brightness in the optical and $\gamma$-ray bands. We present optical photometric and polarimetric data acquired by the Whole Earth Blazar Telescope (WEBT) Collaboration from 2022 April 6 to July 6.
Many episodes of unprecedented fast variability were detected, implying an upper limit to the size of the emitting region as low as  $10^{-4}$ parsec. 
    The WEBT data show rapid variability in both the degree and angle of polarization. We analyse different models to explain the polarization behaviour in the framework of a twisting jet model, which assumes that the long-term trend of the flux is produced by variations in the emitting region viewing angle. All the models can reproduce the average trend of the polarization degree, and can account for its general anticorrelation with the flux, but the dispersion of the data requires the presence of intrinsic mechanisms, such as turbulence, shocks, or magnetic reconnection.
The WEBT optical data are compared to $\gamma$-ray data from the {\it Fermi} satellite. 
These are analysed with both fixed and adaptive binning procedures.
We show that the strong correlation between optical and $\gamma$-ray data without measurable delay assumes different slopes in faint and high brightness states, and this is compatible with a scenario where in faint states we mainly see the imprint of the geometrical effects, while in bright states the synchrotron self-Compton process dominates.
\end{abstract}



\section{Introduction}

With the term ``blazar" we indicate a jetted active galactic nucleus (AGN) with one jet directed towards us.
Leptons moving at relativistic speeds along the magnetic field lines inside the jet produce low-energy synchrotron radiation and high-energy radiation through inverse-Compton scattering of soft photons. Processes involving hadrons may also be responsible for the high-energy emission \citep[e.g.][]{boettcher2013}. 
Because of the jet orientation, this radiation is relativistically Doppler beamed \citep[e.g.][]{urry1995}.
Consequences of the Doppler beaming are that the flux that we observe is enhanced in comparison to what is emitted by the source, and the variability time scales are shortened. This is why blazars often show extreme variability at all wavelengths, from the radio to the $\gamma$ rays, on time scales ranging from years to minutes \citep[e.g.][]{wagner1995,aharonian2007,albert2007,shukla2020,weaver2020}. The origin of such multiscale flux changes is still debated, but it is clear that different processes must intervene to account for the variety of observed variability events. Flares suggest that particles get accelerated in the jet. The two main acceleration mechanisms that are commonly invoked are shock waves propagating in the jet \citep[e.g.][]{hughes1985,marscher1985}, and magnetic reconnection, possibly triggered by kink instabilities \citep[e.g.][]{sironi2015,zhang2018,bodo2021,zhang2022}. Moreover, turbulence is likely to play an important role \citep[][]{marscher2014}.
But since Doppler beaming depends on the viewing angle, strong flux variations are expected if the jet emitting regions change their orientation with respect to the line of sight \citep[e.g.][]{raiteri2017_nature}. This can happen because of jet precession, or rotation induced by orbital motion in a supermassive black hole (BH) binary system, or jet twisting due to kink instabilities developing inside the jet. 

The jet physics is determined by the magnetic field, and this can be studied by means of polarimetric observations. In blazars, both the polarization degree and the polarization angle are very variable \citep[e.g.][and references therein]{raiteri2021_galaxies}. 
Their behaviour is often uncorrelated with the total flux density, which makes the interpretation of the polarization observations very difficult.

The above variability features are common to both the blazar subclasses, i.e.\ flat-spectrum radio quasars (FSRQs) and BL Lacertae-type objects (BL Lacs). The two types have originally been distinguished on the basis of the strength of the emission lines in their spectra \citep{stickel1991,stocke1991}, the former showing broad emission lines with equivalent width  greater than 5\AA\ in the rest frame, the latter exhibiting (nearly) featureless spectra.

S4~0954+65 is a BL Lac-type blazar at redshift $z=0.3694$ \citep{becerra2021}, which is well known for its strong radio and optical flux variability, also on short time scales \citep[e.g.][]{wagner1993,heidt1996,raiteri1999,papadakis2004,marchili2012,bachev2015,raiteri2021}. 
\citet{heidt1996} noted several symmetric optical outbursts and suggested that geometrical effects may play an important role in explaining the source variability.

Polarimetric observations show that the polarization degree undergoes rapid changes at both radio \citep{gabuzda2000} and optical wavelengths \citep{morozova2014,raiteri2021}. Wide rotations of the polarization angle have been observed \citep{hagen2015}.

The source was detected at GeV energies by the Compton Gamma Ray Observatory
\citep[CGRO;][]{mukherjee1995} and at
TeV energies by the Major Atmospheric Gamma Imaging Cherenkov (MAGIC) telescopes during an exceptionally bright optical state in 2015 \citep{magic2018}.

\citet{raiteri2021} analysed well-sampled optical light curves obtained in 2019--2020 by the Whole Earth Blazar Telescope\footnote{\url{https://www.oato.inaf.it/blazars/webt/}} \citep[WEBT; e.g.][]{villata2002,villata2006,raiteri2017_nature,larionov2020,jorstad2022,raiteri2023} Collaboration, with the addition of 2-min cadence data acquired by the Transiting Exoplanet Survey Satellite ({\it TESS}) during three observing sectors of about 1 month each.
They detected several characteristic times scales of variability, ranging from 6 hours in the {\it TESS} light curves, to several weeks in the whole data set.
Moreover, they identified quasi-periodic oscillations (QPOs) with period of about 1 month, which were interpreted as produced by a rotating inhomogeneous helical jet, whose pitch angle changes in time.

A QPO of 1.52~d  was detected in the {\it TESS} light curves
by \citet{kishore2023}. They suggested that the QPO may originate from the orbital motion of some blob or flare in the innermost part of the accretion disc, and estimated a BH mass of $\sim 2 \times 10^8 M_\odot$ or $\sim 10^9 M_\odot$ for a Schwarzschild or Kerr BH, respectively.

QPOs with periods of 66~d and 210~d were recognised by \citet{gong2023} in the $\gamma$-ray light curves from the {\it Fermi} satellite. The most plausible scenario was found to be a plasma blob following a helical path inside the jet.

In April 2022, S4~0954+65 was observed in very high optical \citep{bachev2022, vlasyuk2022a, vlasyuk2022b} and $\gamma$-ray \citep{rani2022} states. This triggered intensified observations by the WEBT \citep{marchini2022}, whose members have been regularly monitoring the source since 2007 through the GLAST-AGILE Support Program \citep[GASP, e.g.][]{villata2008,villata2009}.

In this paper we present the results of the optical monitoring by the WEBT in the period 2022 April 6 to July 6, together with the $\gamma$-ray observations by the LAT instrument onboard the {\it Fermi} satellite. We explore the details of the flux variability at both low and high energies and analyse their correlation. We also investigate the optical polarization behaviour.

\section{Optical photometry}

Optical observations were carried out in the framework of the WEBT Collaboration at the observatories listed in Table~\ref{tab:webt}.

\begin{table*}
\caption{Details on the optical datasets contributing to this paper.}
\label{tab:webt}
\begin{tabular}{llrrcc}
\hline
Dataset                &  Country        & Diameter (cm) & $N_{\rm obs}$ & Symbol & Colour \\
\hline
Abastumani              & Georgia &  70  & 888  & {\LARGE $\diamond$} & dark green \\ 
ARIES                   & India   & 130  & 87   & $\square$ & green \\ 
Athens$^a$              & Greece  &  40  & 177  & {\large $+$} & grey\\
Belogradchik$^b$        & Bulgaria & 60 & 42 & {\large $+$} & cyan\\
Burke-Gaffney           & Canada & 61 & 18 & {\LARGE $\diamond$} & pink\\
Calar Alto$^b$          & Spain & 220 & 2 & {\LARGE $\ast$} & red \\
Catania (SLN)           & Italy & 91 & 3 & {$\triangle$} & blue \\
Connecticut             & US & 51 & 2 & {\LARGE $\ast$} & grey\\
Crimean (ST-7)          & Crimea & 70 & 1  & {\large $+$} & magenta \\
Crimean (ST-7; pol)$^b$ & Crimea & 70 & 96 & {$\times$} & dark green \\
Hans Haffner         & Germany & 50 & 194 & {\LARGE $\circ$} & red \\
Lulin (SLT)          & Taiwan  & 40 & 38 & {\LARGE $\circ$} & black \\
McDonald (LCO)       & US & 40 & 1 & {$\times$} & black\\
Mt. Maidanak         & Uzbekistan & 60 & 101 & {\LARGE $\circ$} & violet \\
Osaka Kyoiku         & Japan & 51 & 276 & {$\square$} & orange  \\
Rozhen               & Bulgaria & 200 & 6  & {$\square$} & red \\
Rozhen               & Bulgaria & 50/70 & 10  & {$\times$} & orange \\
San Pedro Martir     & Mexico & 84 & 8 & {$\square$} & blue\\
SAO RAS              & Russia & 100 & 43 & {\LARGE $\circ$} & blue\\
SAO RAS              & Russia &  50 & 706 & {\LARGE $\diamond$} & red\\
Siena                & Italy & 30 &  438 & {\LARGE $\diamond$} & blue\\
Skinakas (RoboPol)$^b$  & Greece & 130 & 16 & {$\times$} & blue\\
St. Petersburg$^b$   & Russia & 40 & 53 & {\large $+$} & orange\\
Teide (IAC80)        & Spain & 80 & 6 & {\LARGE $\ast$} & green\\
Teide (LCO)          & Spain & 100 & 1 & {\large $\ast$} & black\\
Teide (LCO)          & Spain & 40 & 9 & {\large $+$} & black\\
Tijarafe             & Spain & 40 & 130 & {$\triangle$} & green \\
Valle d'Aosta        & Italy &  80 & 10 & {\large $+$} & violet\\
Vidojevica$^c$       & Serbia & 140 & 64 & {$\square$} & black\\
Vidojevica$^c$       & Serbia & 60 & 12 & {$\triangle$} & black\\
West Mountain        & US & 91 & 190  & {\LARGE $\circ$} & dark green\\
\hline
\end{tabular}\\
``LCO" refers to telescopes belonging to the Las Cumbres Observatory global telescope network\\
$^a$ University of Athens Observatory (UOAO)\\
$^b$ Also polarimetry\\
$^c$ Astronomical Station Vidojevica\\
\end{table*}

\begin{figure*}
 \includegraphics[width=12cm]{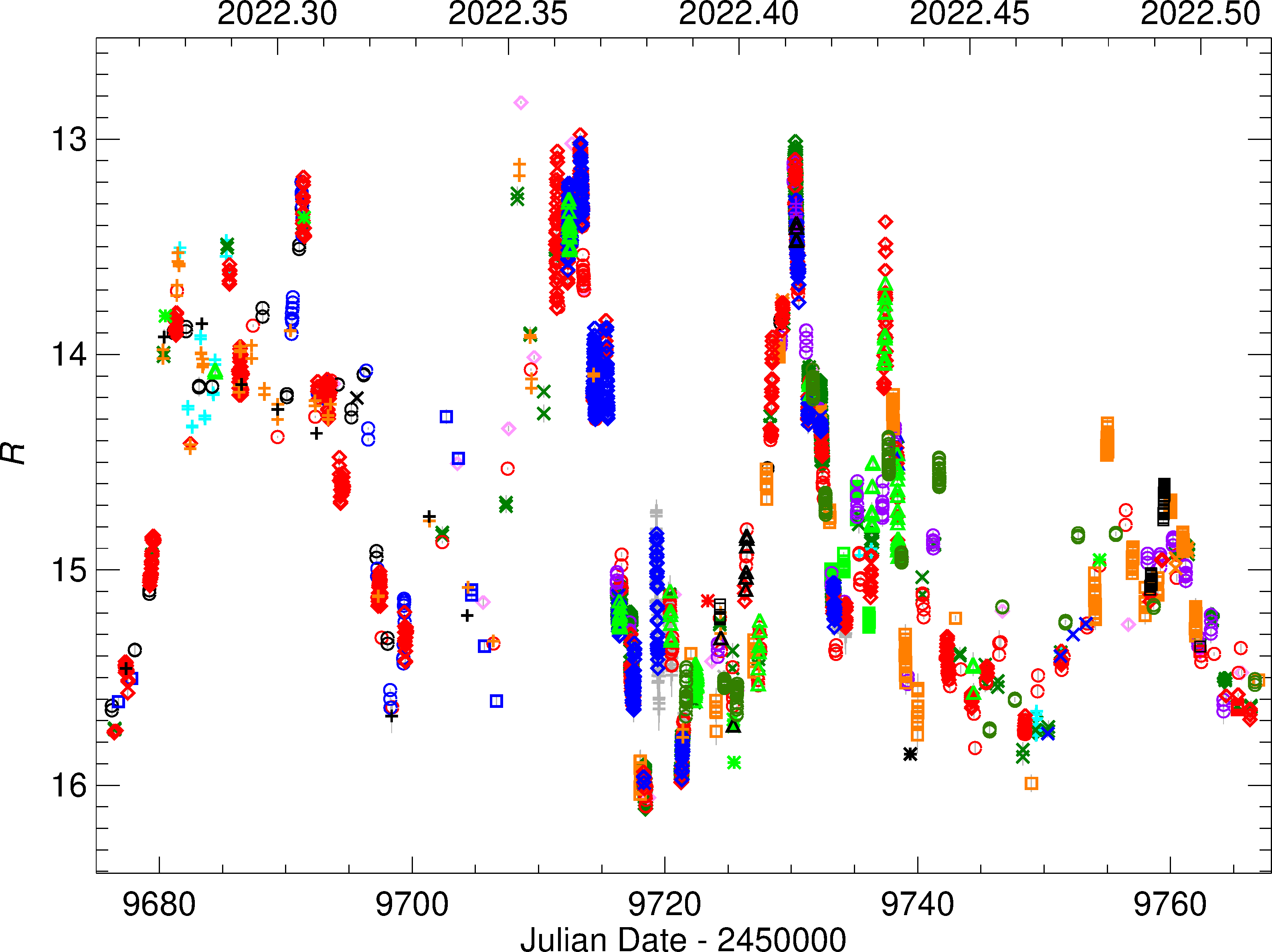}
 \caption{The $R$-band optical light curve of S4~0954+65 in the 2022 flaring period analysed in this paper. The various datasets are distinguished by different colours and symbols as specified in Table \ref{tab:webt}. Uncertainties are plotted in grey and are typically smaller than the symbol size.}
 \label{fig:webt}
\end{figure*}

The source magnitude was derived using the photometric sequence published by \citet{raiteri1999}. Data belonging to 31 datasets from 25 observatories in 16 countries around the northern hemisphere were carefully assembled and processed to get a homogeneous and precise optical light curve in $R$ band. No significant systematic offset of the data points of individual datasets was found with respect to the others. Particularly noisy datasets from the same telescope were binned over time intervals of a few minutes and clear outliers were eliminated. The final light curve is shown in Fig.~\ref{fig:webt}; it includes 3628 data points acquired during 91 days, from April 6 to July 6 2022 ($\rm JD=2459676.0$--2459767.0).
In this period, the source showed wide brightness variations, with a maximum amplitude of 3.28 mag, and reached its maximum brightness level $R=12.83 \pm 0.01$, exceeding the levels observed during the 2015 outburst \citep{bachev2015,magic2018}.

\section{Intra-day variability}
As mentioned in the Introduction, S4~0954+65 has often shown intense intra-day variability. One of the most extreme episodes was reported by \citet{bachev2015}, who observed an almost 0.7 mag brightness decrease in 5 h.
In the period analysed here, we found even more dramatic fast variability. 

The brightness rise that brought the source to its historical maximum $R=12.83 \pm 0.01$ on May 9 ($\rm JD \approx 2459708.61$) involved a variation of roughly 2.8 mag in less than 2~d (47~h).
Three days later we observed a 0.73 mag brightness increase in 1.2~h.

\begin{figure}
\includegraphics[width=\columnwidth]{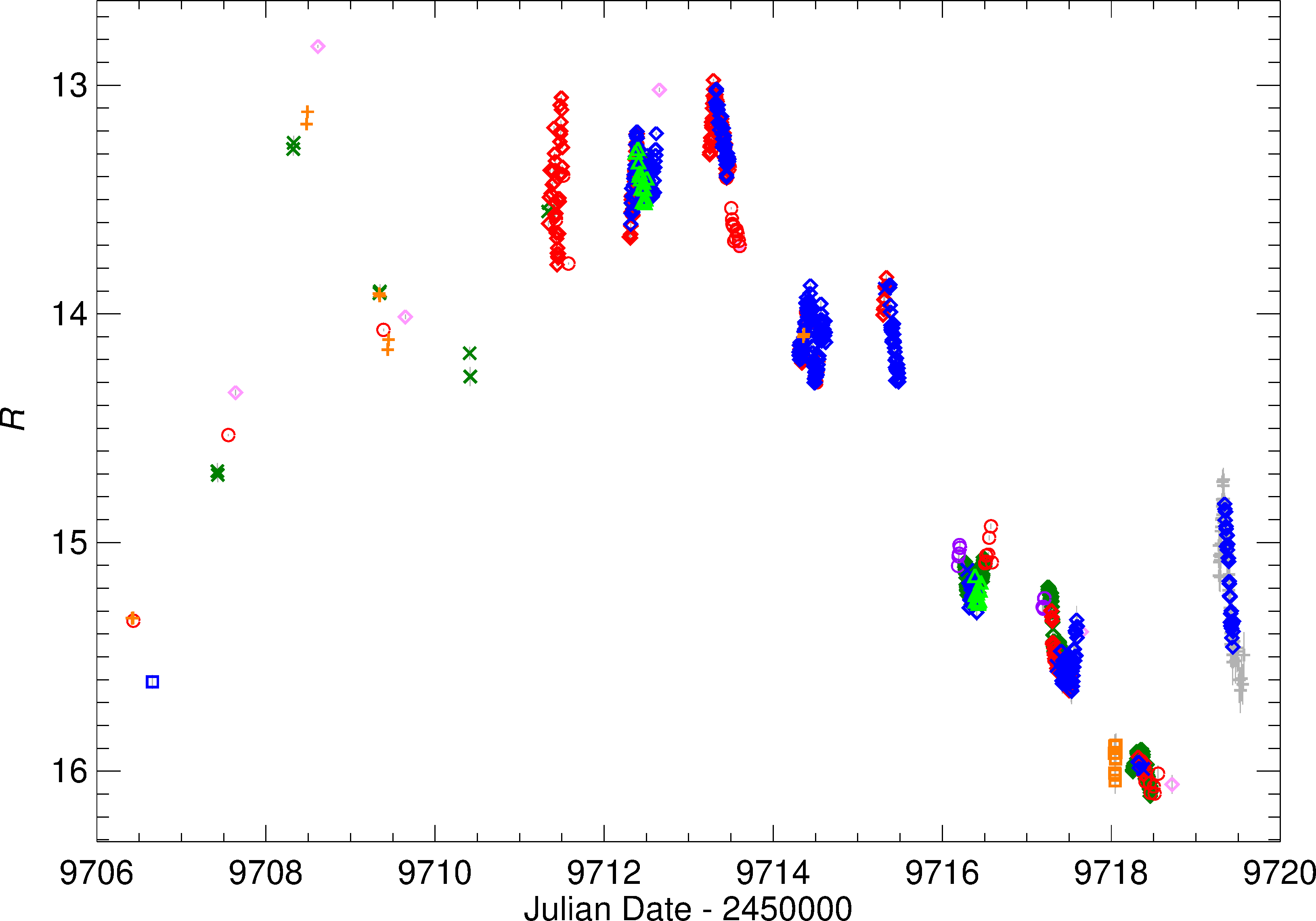}
    \caption{An enlargement of the $R$-band light curve during the extreme variability phase observed in May 6--20, 2022.}
    \label{fig:idv1}
\end{figure}

Another extreme IDV episode was detected on May 19--20 (JD 2459719), with a $\sim 0.92 \, \rm mag$ fading in 4.8~h.
Fig.~\ref{fig:idv1} displays the source light curve in this period.

\begin{figure}
\includegraphics[width=\columnwidth]{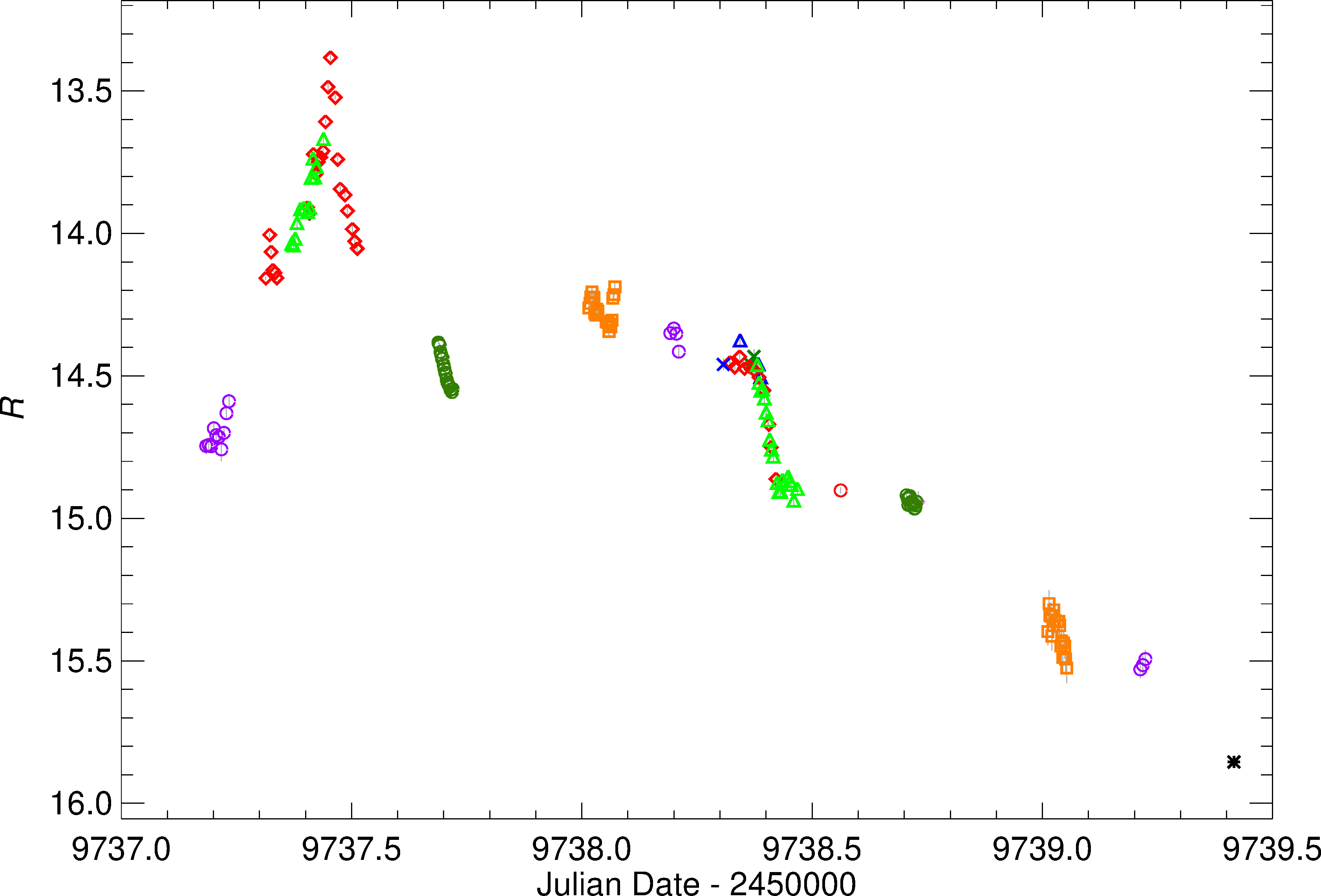}
    \caption{An enlargement of the $R$-band light curve during the extreme variability phase observed in June 6--8, 2022.}
    \label{fig:idv2}
\end{figure}

On June 6--7 (JD 2459737) we observed a $\sim 1.4 \, \rm mag$ brightness increase in about 5.7~h, followed by a $\sim 0.67 \, \rm mag$ brightness decrease in about 1.4~h, which was part of a longer dimming phase of $\sim 1.2 \, \rm mag$ in less than 10~h. This fading trend continued in the following two days, with a total variability amplitude of about 2.5 mag in less than 2~d (47~h). An enlargement of the optical light curve in this period is shown in Fig.~\ref{fig:idv2}.

Overall, on 17 occasions we detected mag changes of more than 0.5 mag in less than 12~h.

\section{Size of the optical emitting region}

\begin{figure}
\includegraphics[width=\columnwidth]{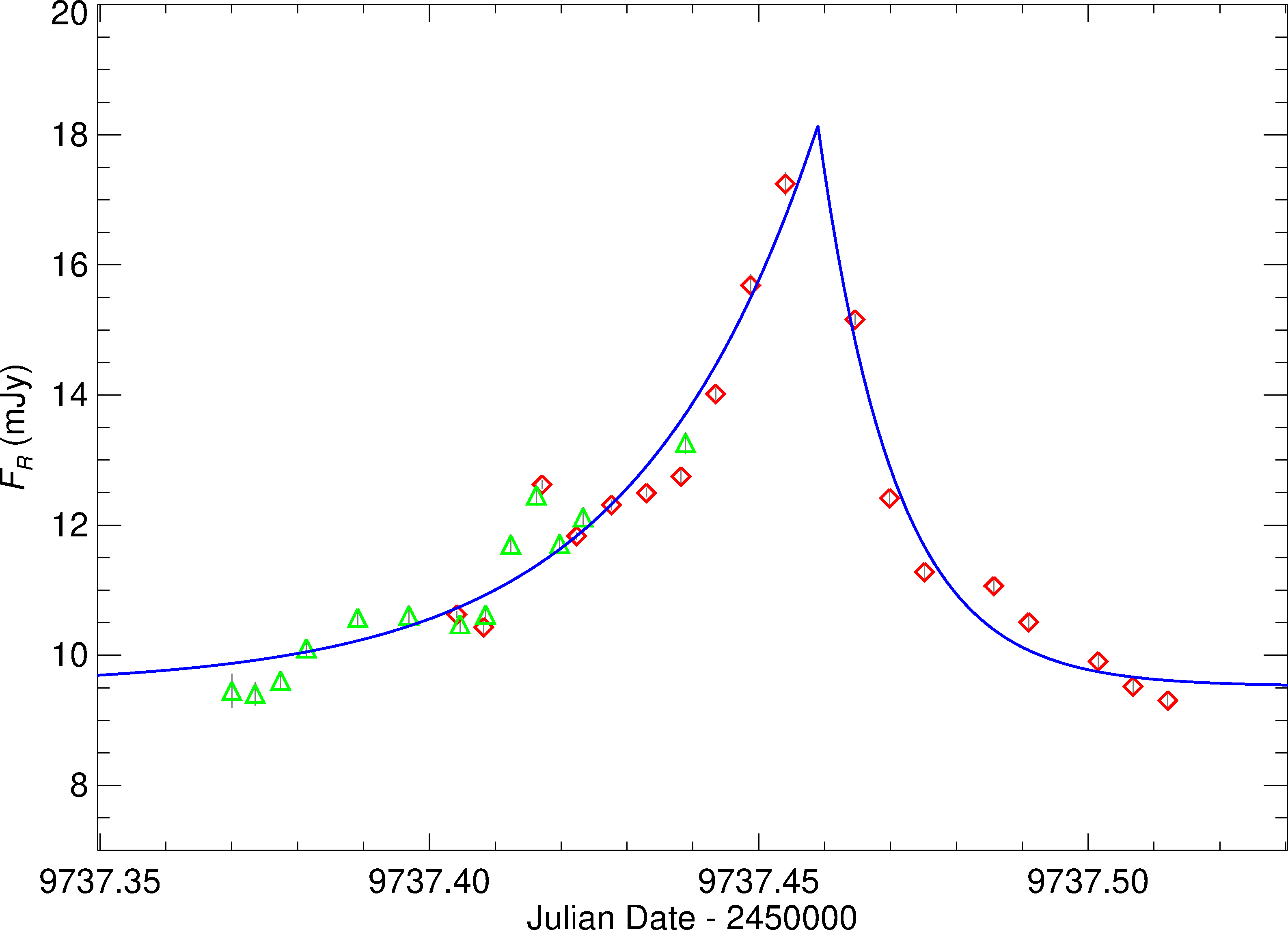}
    \caption{The rapid flare on $\rm JD=2459737$ fitted by the model in Eq.~\ref{eq:valta}. }
    \label{fig:mf}
\end{figure}

As detailed in the previous section, unprecedented intraday variability was observed in the period analysed.
From causality arguments based on light travel time, the observed minimum variability time scale $\Delta t_{\rm min}$ can put an upper limit to the size $R$ of the emitting region:
\begin{equation}
    R < c \, \Delta t_{\rm min} \times \delta/(1+z)
\end{equation}
where $c$ is the speed of light, $\delta$ is the Doppler factor and $z$ the source redshift.
The value of $\Delta t_{\rm min}$ can be obtained 
from the well-sampled, extreme IDV episode on JD 2459737, which is shown in Fig.~\ref{fig:mf}. Flux densities have been obtained from magnitudes using the calibrations by \citet{bessell1998} and correcting for Galactic absorption ($A_V=0.259 \, \rm mag$ from the NASA/IPAC Extragalactic Database\footnote{\url{https://ned.ipac.caltech.edu/}}).

We modelled the flare according to \citet{valtaoja1999}:
\begin{equation}
\begin{split}
    F  = F_0+A e^{(t-t_{\rm peak})/\Delta t_1} \quad \textrm{if} \quad  t< t_{\rm peak}\\
    F = F_0+A e^{(t_{\rm peak}-t)/\Delta t_2} \quad \textrm{if} \quad  t \geq t_{\rm peak}\\
    \end{split}
    \label{eq:valta}
\end{equation}
where $F_0$ is the base level, $A$ the flare amplitude, $t_{\rm peak}$ the time of the flare peak, and $\Delta t_1$ and $\Delta t_2$ the time scales before and after the peak, respectively.

The least-squares best-fit model of the flare is obtained by the following parameters: 
$F_0=(9.5 \pm 0.5) \, \rm mJy$, $A= (8.6 \pm 1.1) \, \rm mJy$, 
$t_{\rm peak} = (2459737.459 \pm 0.003)$, 
$\Delta t_1= (0.028 \pm 0.007) \, \rm d$, and $\Delta t_2=(0.012 \pm 0.005) \, \rm d$.
Setting $\Delta t_{\rm min} \approx 17 \, \rm  min$, and $\delta=13.6$ (see Section~\ref{sec:doppler} and Fig.~\ref{fig:doppler}) we found 
$R < 3 \times 10^{14} \rm \, cm$, i.e.~ about $10^{-4}$ parsec.
This upper limit to the emitting region size 
responsible for the flare is in general smaller than the typical size assumed for the blazar jets, and in particular it is more than 66 times smaller than that assumed by \citet{raiteri1999} when applying the homogeneous model by \citet{ghisellini1998} to the SEDs of S4~0954+65 during a faint state observed in 1994--1998. 
This suggests that we may be seeing flux fluctuations in a jet subregion.

We note that blazar microvariability with as short as a few minutes time scale was also detected at $\gamma$ rays, in both GeV and TeV energy domains. In the case of 3C~279 observed by the {\it Fermi} satellite, a very fast flare in 2018 was ascribed to magnetic reconnection in a region of about $8 \times 10^{14} \rm \, cm$ \citep{shukla2020}. For the microvariations observed by the High Energy Stereoscopic System (H.E.S.S.) in PKS~2155-304 \citep{aharonian2007}, and by MAGIC in Mkn~501 \citep{albert2007}, the inferred sizes are likely more than 10 times smaller, when typical values of $\delta \sim 10$ are assumed.

\section{Optical polarimetry}
\label{sec:pola}
\begin{figure}
\includegraphics[width=\columnwidth]{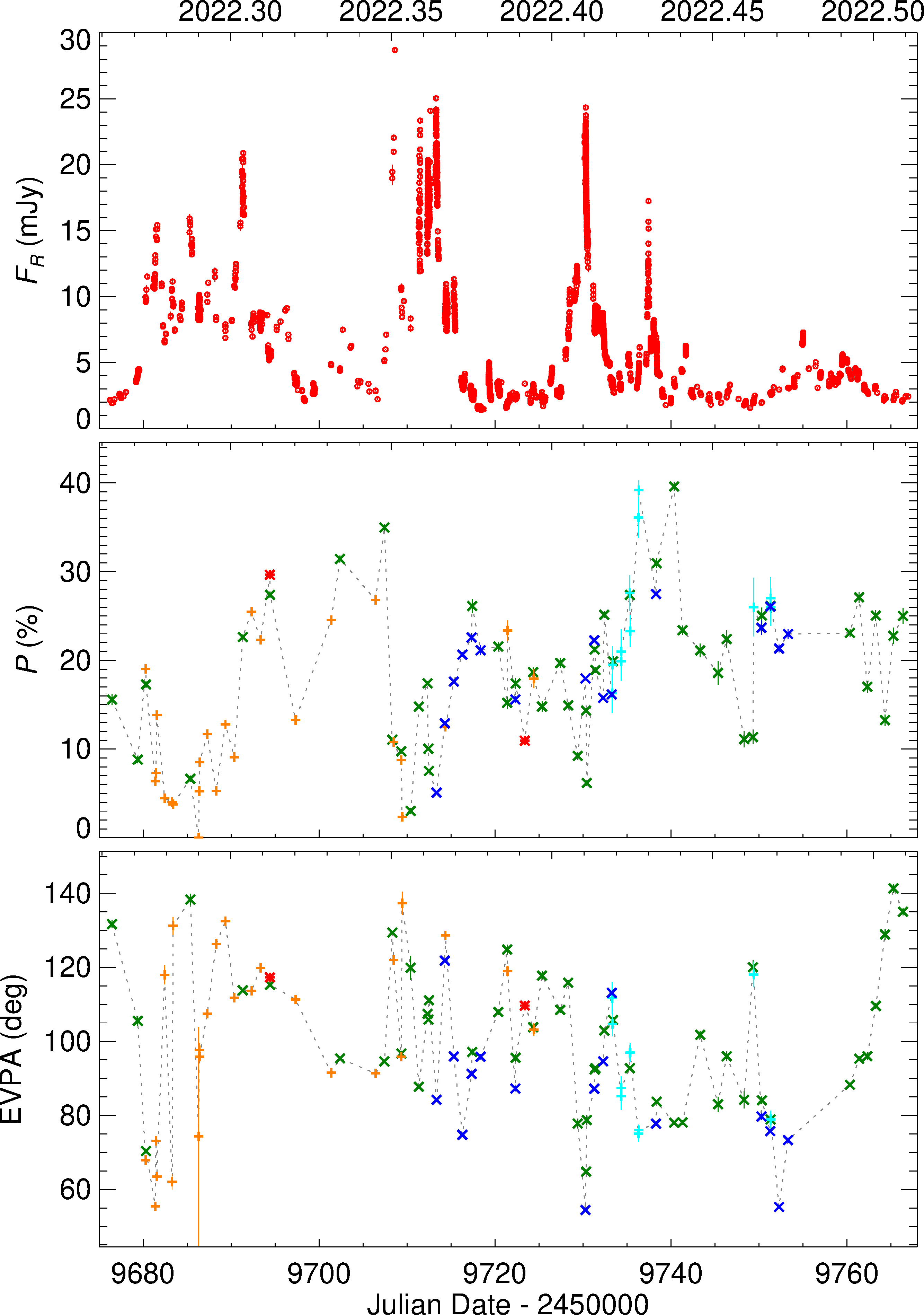}
    \caption{Top: De-reddened optical flux densities in the $R$ band. Middle: Polarisation degree. Bottom: Polarisation angle. Different symbols and colours distinguish different contributions, as listed in Table~\ref{tab:webt}.}
    \label{fig:pola}
\end{figure}

Optical polarimetric data for this work were acquired at the Belogradchik, Calar Alto, Crimean, Skinakas, and St.~Petersburg observatories.
The degree of polarization $P$ and the electric vector position angle (EVPA) are shown in Fig.~\ref{fig:pola} together with the optical flux densities.

\begin{figure}
\includegraphics[width=\columnwidth]{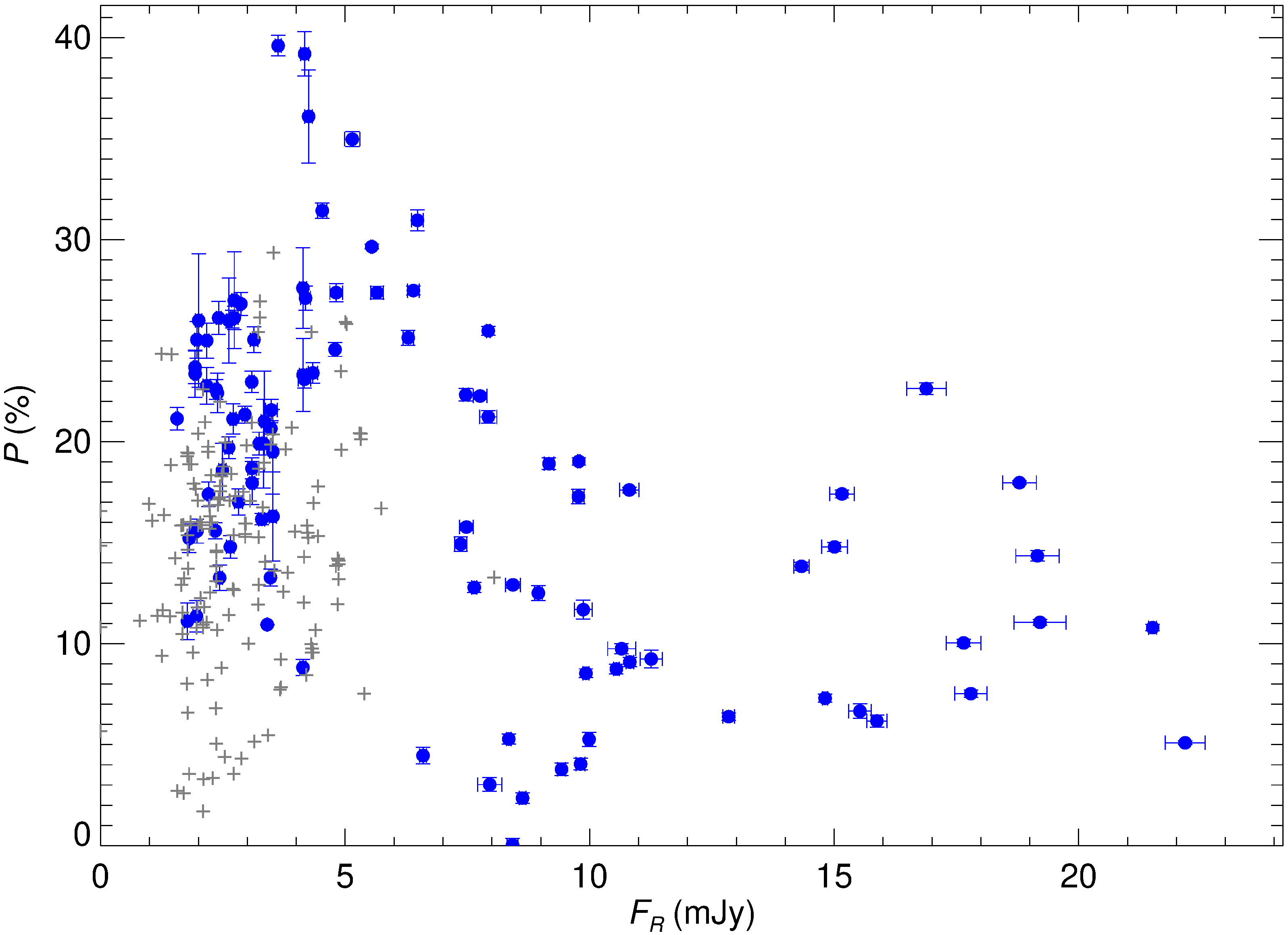}
    \caption{Polarisation degree versus optical flux density in the $R$ band. Blue dots refer to data taken in the period covered by this paper; grey plus signs to those acquired in 2019--2020 and published in \citet{raiteri2021}.}
    \label{fig:pf}
\end{figure}

Both $P$ and EVPA display a flickering behaviour.
The values of $P$ range from 0.03\% to 39.6\%, those of EVPA span a $\sim 87 \degr$ interval.
A general anticorrelation between $P$ and the flux density in $R$ band, $F_R$, is recognisable, which is confirmed by the plot of $P$ versus $F_R$ shown in Fig.~\ref{fig:pf}. Values of $P$ higher than 23\% are found only when $F_R<8 \, \rm mJy$, and the two values greater than 39\% correspond to flux densities as low as $\sim 4$ mJy. We note that values of $P$ of $\sim 40\%$ are close to the maximum values observed in blazars. Here they are found just before and just after the very narrow flare on JD 2459737 that seems to conclude the preceding period of strong activity. 

\begin{figure}
\includegraphics[width=\columnwidth]{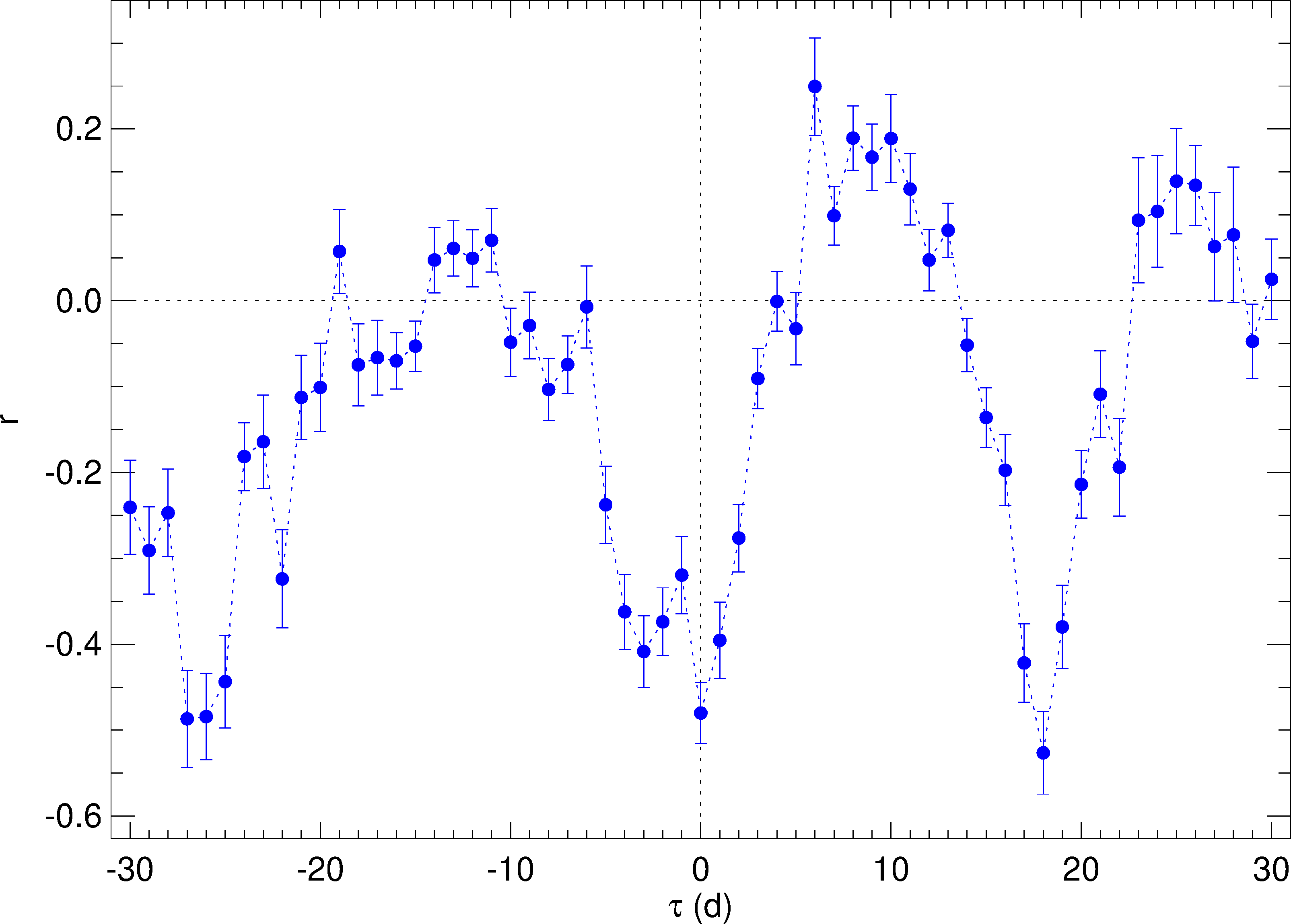}
    \caption{DCF between the optical flux densities $F_R$ and the polarization degree $P$.}
    \label{fig:dcf_op}
\end{figure}

Another way of assessing the general anticorrelation between $P$ and $F_R$ is through the discrete correlation function \citep[DCF;][]{edelson1988,hufnagel1992}. A strong correlation results in a positive DCF peak with value close to one, while an anticorrelation gives negative DCF values. As visible in Fig.~\ref{fig:dcf_op}, the DCF between $F_R$ and $P$ assumes negative values for time lags around zero and in general does not show positive peaks greater than 0.25.

It is interesting to compare the source polarization behaviour in 2022 with that observed in 2019--2020 and described by \citet{raiteri2021}. At that time, the source flux density was lower, oscillating between $F_R \sim 1 \rm \, mJy$ and $\sim 10 \rm \, mJy$, while $P$ was ranging from about 2\% to 29\%. 
The authors commented that $P$ did not seem to correlate with flux, thus the general anticorrelation seen in the current dataset was previously not detected. We plotted these earlier data in Fig.~\ref{fig:pf}. We first note that in the common range of flux densities, the values of $P$ in 2019--2020 were in average lower than in 2022, which could be due to a more turbulent magnetic field in that period.
Then we infer that the anticorrelation was not detected at that time because it shows up clearly only when the source exceeds the brightness values that were observed in 2019--2020.

\section{Interpretation of the flux and polarization behaviour}
\label{sec:doppler}

In several papers by the WEBT Collaboration we have proposed that the long-term flux variability at a given frequency is due to orientation changes of the corresponding jet emitting region, which produce a variation of the Doppler beaming. 
In contrast, the fast variability superimposed on the long-term trend is likely produced by intrinsic processes, such as turbulence, shocks or magnetic reconnection.

\begin{figure}
\includegraphics[width=\columnwidth]{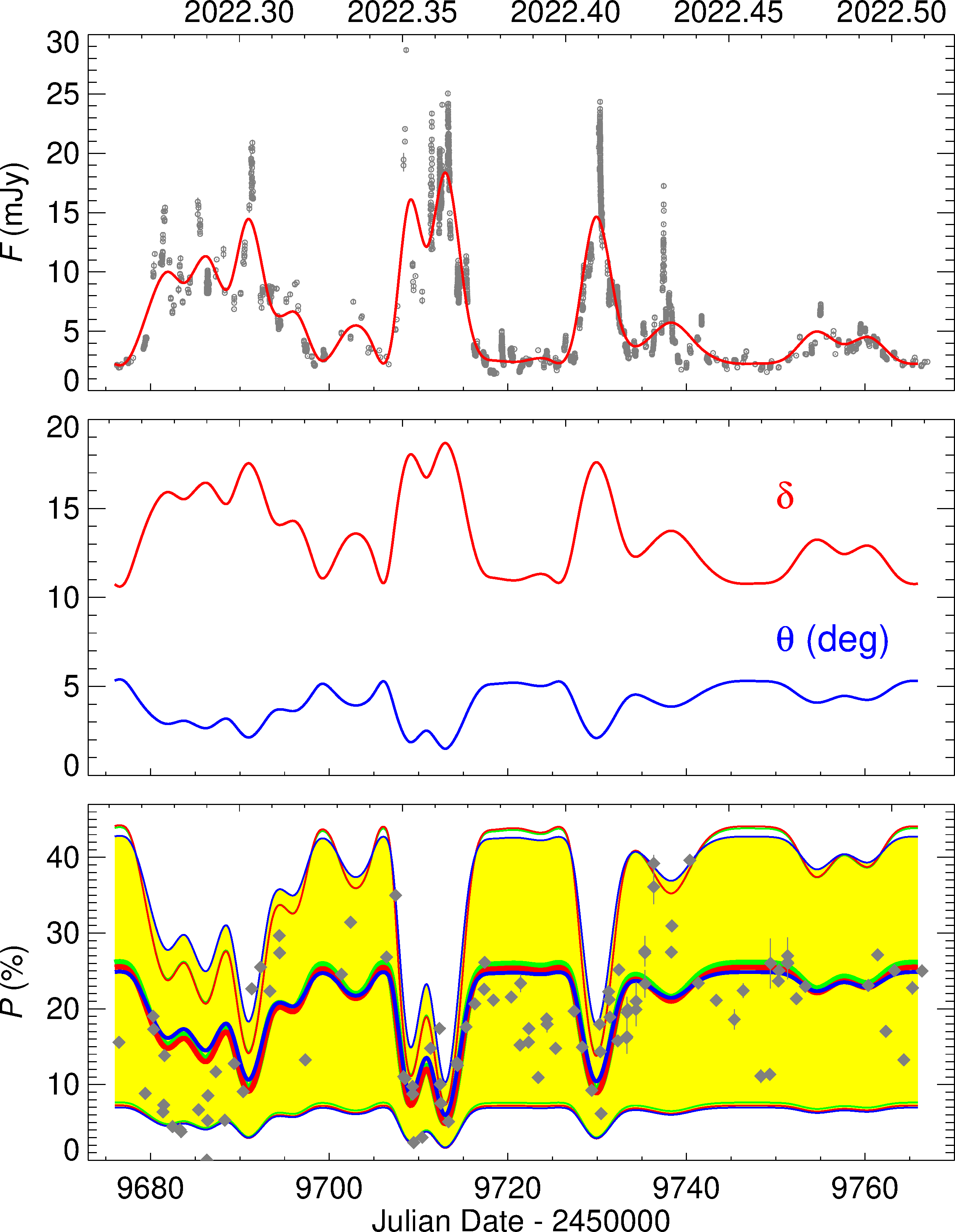}
    \caption{Top: $R$-band flux densities corrected for Galactic extinction (grey dots); the red line represents the long-term trend obtained as a cubic spline interpolation through the light curve binned over time intervals depending on brightness.
    Middle: Behaviour of the Doppler factor $\delta$ (red line) and viewing angle $\theta$ (blue line) in time derived from the long-term trend.
    Bottom: Observed degree of polarization (grey diamonds); the blue, red, and green thick lines represent predictions by the models in Eq.~\ref{lyu} with $P_{\rm max}=25\%$, Eq.~\ref{par} with $\Omega=7$, and Eq.~\ref{hug} with $\eta=1.4$, respectively. The yellow area is delimited by the  models in Eq.~\ref{lyu} with  $P_{\rm max}$ between 7\% and 43\%, which are very similar to those by Eq.~\ref{par} with $\Omega=5.4$--9.7, and to those by Eq.~\ref{hug} with $\eta=1.10$--1.85.}
    \label{fig:doppler}
\end{figure}

We can obtain the optical long-term trend by performing a cubic spline interpolation through the light curve shown in Fig.~\ref{fig:pola}, after binning it over variable time intervals that decrease with increasing brightness in order to take into account the effect of Doppler beaming on time scales \citep[see discussion in][]{raiteri2017_nature}.
The long-term trend derived in this way is shown in Fig.~\ref{fig:doppler}, and in our view it represents the amount of variability that can be ascribed to changes of the viewing angle. Because the flux density depends on the Doppler factor $\delta$ as
$F_\nu \propto \delta^{n+\alpha}$,
where $n=2$ for a continuous jet \citep{urry1995} and $\alpha=1.8$ is the source mean optical spectral index \citep{raiteri2021}, from the long-term trend of the flux (i.e.\ the spline) we can derive the behaviour in time of $\delta$, which is shown in Fig.~\ref{fig:doppler}.
Moreover, we know that $\delta$
depends on the viewing angle $\theta$ as:
$\delta=[\Gamma \, (1-\beta \, \cos \theta)]^{-1}$, where $\Gamma=(1-\beta^2)^{-1/2}$ is the bulk Lorentz factor and $\beta$ is the plasma bulk velocity in units of the speed of light. Therefore, we can also infer the trend of $\theta$ in time, once reasonable assumptions are made on the values of the other parameters.
The trends of $\delta$ and $\theta$ shown in Fig.~\ref{fig:doppler} were obtained by fixing
$\Gamma=10$ and $\theta_{\min}=1.5 \degr$.
The first value is the same adopted by \citet{raiteri2021} and is comparable with that obtained by \citet{jorstad2017}; in contrast, $\theta_{\min}$ was halved with respect to that in \citet{raiteri2021} because the brightness levels in the current dataset are much higher.
Indeed, at this minimum viewing angle, the Doppler factor reaches its maximum value $\delta \approx 18.7$, and consequently we observe the maximum brightness level of the long-term trend. In contrast, the lowest value of $\delta \approx 10.6$ corresponds to the largest value of $\theta \approx 5.4 \degr$ and to the faintest long-term flux level.

What is the prediction of this geometrical scenario for the behaviour of the polarization? We investigated different possibilities.
\citet{lyutikov2005} analysed different models of relativistic jets characterised by helical magnetic fields.
The authors assume a cylindrical shape for the jet, with the emitting plasma moving parallel to the jet axis. They do not consider bulk rotation of the jet.
In cases where the number density of relativistic particles scales with the square of the intrinsic
magnetic field, it is possible to infer \citep[see also][]{raiteri2013} that
\begin{equation}
    P=P_{\rm max} \, \sin^2 \, \theta'
    \label{lyu}
\end{equation}
where $\sin \theta'=\delta \sin \theta$.
The values of $P_{\rm max}$ can be derived from the comparison with the observational data.
In Fig.~\ref{fig:doppler} we report the results obtained with $P_{\rm max}=7, 25$, and 43\%. The case with $P_{\rm max}=25\%$ fairly reproduces the average observed polarization level. However, the data points show stronger variability. There are phases where the observed $P$ is lower than predicted, suggesting that the magnetic field was less ordered. This occurs in particular during the first stage of activity, 
before the flare at $\rm JD \approx 2459690$, and also before the flare at $\rm JD \approx 2459730$. In contrast, there are other phases where $P$ reaches quite high values, as just after these two flares.
In this case, something must have happened in the jet to order the magnetic field.
We note that the observed values of $P$ are roughly included between the models with $P_{\rm max}=7\%$ and $P_0=43\%$.

We next considered the model of a rotating relativistic jet with helical magnetic field by \citet{pariev2003}. 
The jet is cylindrical and the magnetic field has a uniform poloidal component, while the toroidal component decreases from the centre outwards, until it disappears at the jet boundary.
When assuming a value $\zeta=3$ for the index of the power-law energy distribution of the emitting particles, the authors could derive an analytical solution:
\begin{equation}
    P= {{3} \over {4}} {{\left({{\Omega}^2 \over {24}} -1 \right) \sin^2 \theta'} \over {\sin^2 \theta'+ {{\Omega^2} \over {12}} \left(1-{{1} \over {2}} \sin^2 \theta' \right)}}
    \label{par}
\end{equation}
where $\Omega$ is a dimensionless parameter defining 
the angular rotational velocity of the magnetic field lines. Again, we derive the values of $\Omega$ from the data.
As shown in Fig.~\ref{fig:doppler}, values of $\Omega=5.4, 7$, and 9.7 produce a trend of $P$ that is almost equal to that obtained with Eq.~\ref{lyu} and $P_{\rm max}=7, 25$, and 43\%.
The variation of $\Omega$ can come either from an effective change in the angular rotational velocity of the jet or possibly in its radial dependence inside the jet.

Finally, we tried the shock-in-jet model \citep[][see also \citealt{larionov2013} and \citealt{raiteri2013}]{hughes1985}, where a random magnetic field is compressed by the passage of a shock wave. In this case:
\begin{equation}
    P \approx P_0 \,{{(1-\eta^{-2}) \sin^2\theta'} \over {2-(1-\eta^{-2}) \sin^2 \theta'}}
    \label{hug}
\end{equation}
 where $P_0=(\alpha+1)/(\alpha+5/3)$ is the degree of linear polarization for particles with a power-law energy distribution with index $p=2 \, \alpha+1$, and the parameter $\eta$ represents the degree of compression of the magnetic field by the shock wave. As in the previous cases, $\eta$ is set by comparison with the data.
If we set $\eta=1.4$,
we find the same average behaviour as in the two previous models (see Fig.~\ref{fig:doppler}), while by setting  $\eta= 1.10$ and $1.85$ we can reproduce fairly well the lower and upper bounds, respectively. 

In summary, the three models considered above lead to the same results for reasonable choices of their parameters. They all imply
a dependence of the polarization degree on the viewing angle, which anticorrelates with the Doppler factor and thus with the flux, and therefore can explain the general anticorrelation between $P$ and $F_R$ seen in Section \ref{sec:pola}.

\section{$\gamma$-ray observations}
The {\it Fermi} satellite has been monitoring the $\gamma$-ray sky since 2008. 
Light curves produced with data from its Large Area Telescope (LAT) instrument \citep{atwood2009} with 3-day, 1-week, and 1-month binning can be retrieved from the {\it Fermi} LAT Light Curve Repository\footnote{\url{ https://fermi.gsfc.nasa.gov/ssc/data/access/lat/LightCurveRepository/about.html}} \citep{abdollahi2023}. 
\begin{figure}
	\includegraphics[width=\columnwidth]{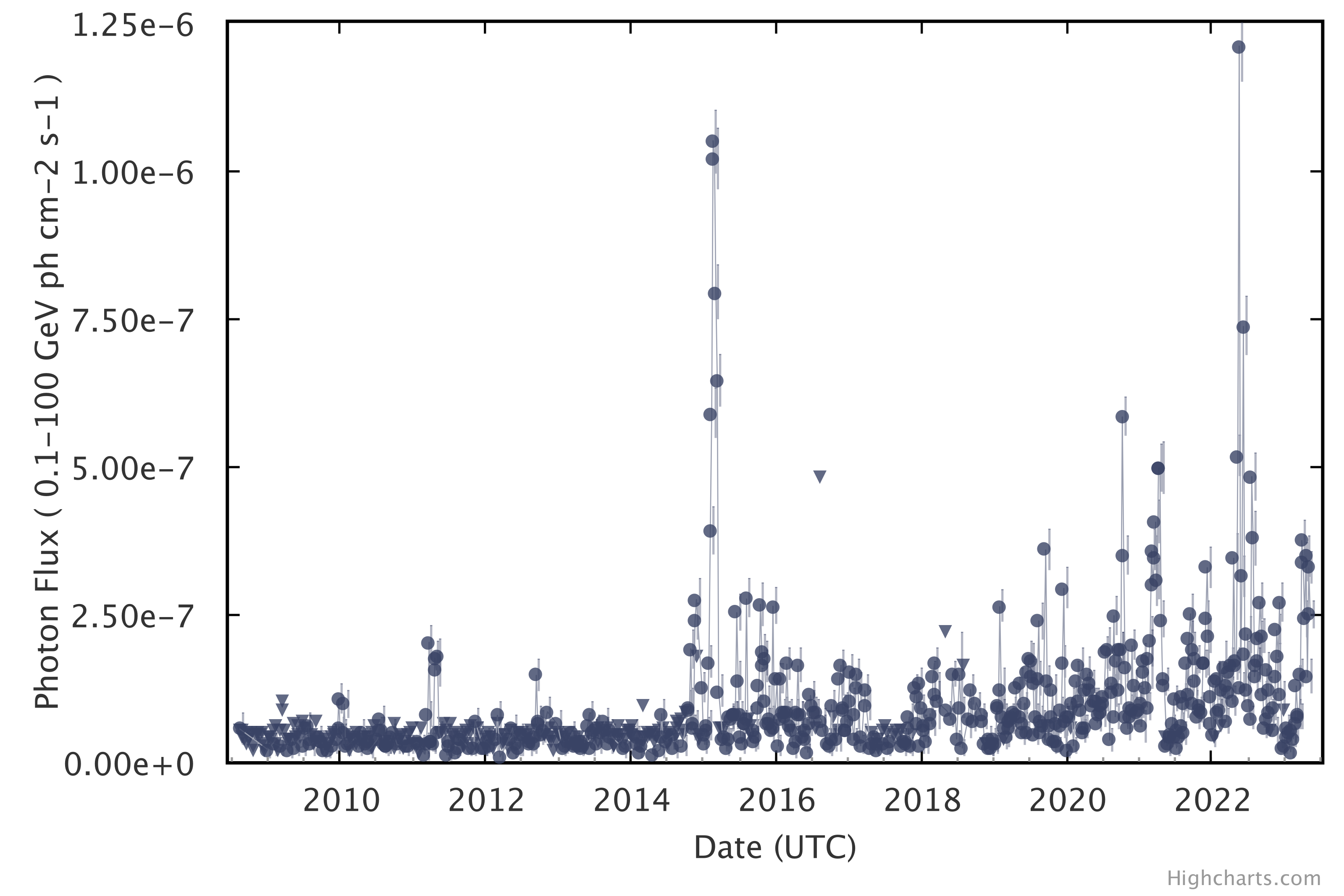}
    \caption{The historical $\gamma$-ray light curve of S4~0954+65 with 1-week binning retrieved from the {\it Fermi} LAT Light Curve Repository.}
    \label{fig:fermi}
\end{figure}
We show the complete 1-week binned light curve of S4~0954+65 in Fig.~\ref{fig:fermi} to put our analysis into context. With this time resolution, the source $\gamma$-ray flux exceeded the $10^{-6} \rm \, ph \, cm^{-2} \, s^{-1}$ level only twice, at the beginning of 2015 and in 2022, and in the latter period it reached the historical maximum. This is the period we are considering in the present paper.
Given the high number of $\gamma$-ray photons detected, we expect to be able to obtain a $\gamma$-ray light curve with a time resolution better than those available in the Repository, for an optimal comparison with the densely sampled optical light curve.
Therefore, we first built $\gamma$-ray light curves with fixed integration time intervals, starting from 5 d and then decreasing the time bin down to 1 h.
By combining the data from these light curves, we could build a composite light curve, with better sampling during the source high states. 
We checked for spectral variations, which were found to be negligible. 
Then we adopted an adaptive binning method with constant spectral shape, which showed to be superior than the fixed binning method from the sampling point of view. Details on the two methods are given below. We believe that the comparison between their results can be instructive.

\subsection{Fixed binning method}

\begin{figure*}
 \includegraphics[width=12cm]{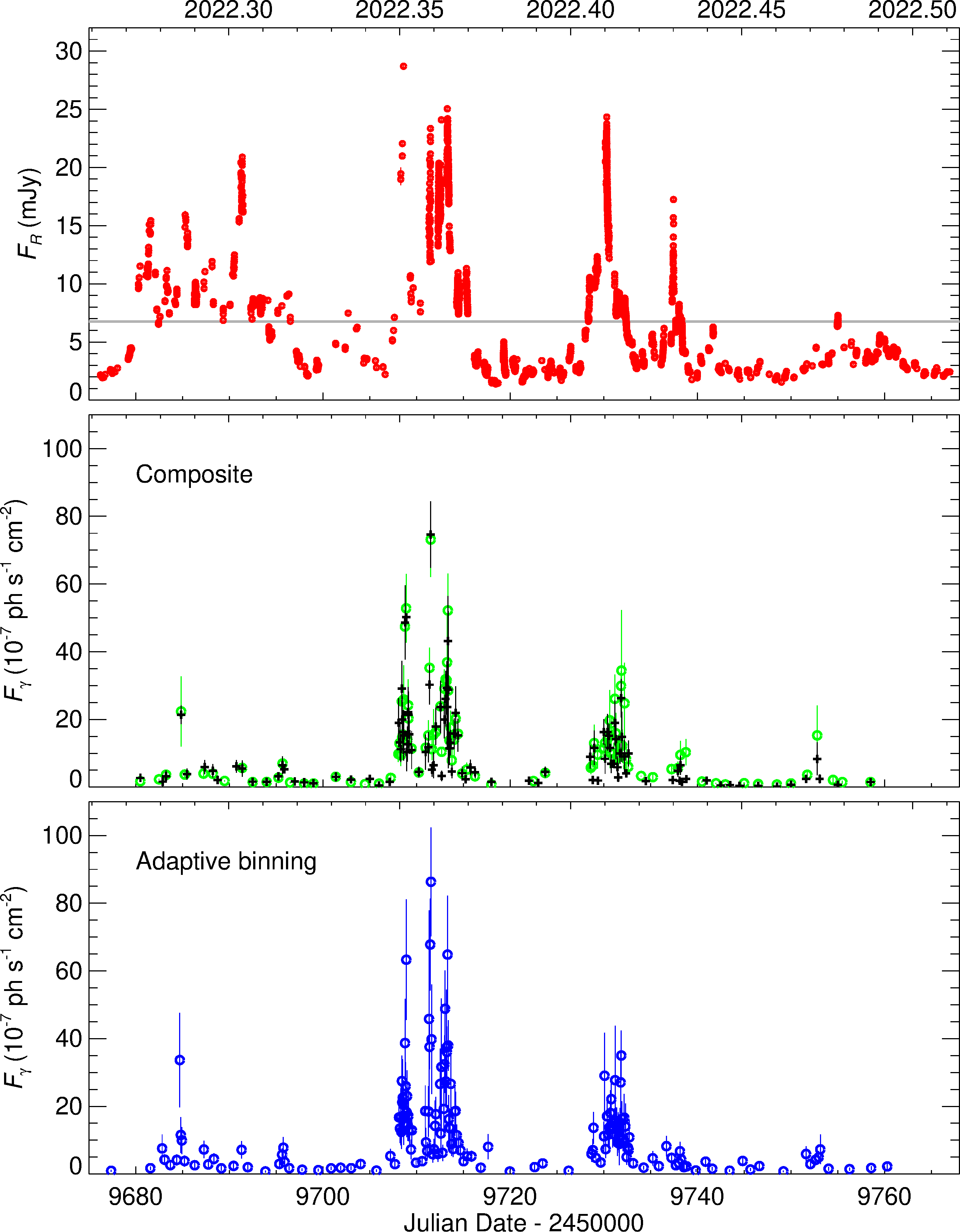}
 \caption{Top: Optical flux densities (mJy).
 The grey horizontal line marks the level $\log \nu F_\nu  =-10.5$ mentioned in Sect.~\ref{sec:gamo}. Middle: Composite $\gamma$-ray light curves; green circles refer to the case with the values of the spectral parameters fixed to $\alpha_{\rm wp}$ and $\beta_{\rm wp}$; black plus signs to the case where they were left free to vary inside default boundaries. Bottom: $\gamma$-ray fluxes obtained with the adaptive binning method.}
 \label{fig:gamop}
\end{figure*}

The $\gamma$-ray data in the time period considered in this paper were analysed in the 0.1--300 GeV energy range using the \texttt{FermiTools} package version 2.2 installed with \texttt{Conda}\footnote{\url{https://github.com/fermi-lat/Fermitools-conda/}}, with instrument response function \texttt{P8R3\_V3}, Galactic diffuse emission model \texttt{gll\_iem\_v07}, and isotropic background model \texttt{iso\_P8R3\_ SOURCE\_V3\_v1}. We performed a binned likelihood analysis, adopting a region of interest of radius 30\degr, a maximum zenith angle of 90\degr\ and only ‘Source’ class events (evclass=128, evtype=3). Fluxes of the sources within a 10\degr\ radius were set as free parameters of the model, whereas fluxes of more distant sources were fixed to their mean values according to the 4FGL catalogue. As in the 4FGL catalogue, for S4~0954+65 (4FGL J0958.7+6534) we used a  log-parabola model of the type:
\begin{equation}
 {\rm d} N/{\rm d} E = N_0 \, (E/E_{\rm b})^{-[\alpha+\beta \log(E/E_{\rm b})] }  
 \label{eq:logparabola}
\end{equation}
with a ``break" energy $E_{\rm b} = 674.472 \, \rm MeV$ and normalisation $N_0$. 
When integrating over the whole period, the maximum likelihood  gives a test statistic $\rm TS = 4426.11$, with an integrated average flux of $(2.78 \pm 0.025) \times 10^{-7} \, \rm ph \, cm^{-2} \, s^{-1}$ and spectral parameters $\alpha_{\rm wp} = 2.033$ and $\beta_{\rm wp} = 0.060$, very close 
to their catalogue values. We calculated light curves with fixed integration time bins of 5, 4, 3, 2, 1 days and 12, 6, 3, 1 hours and  spectral parameters fixed to $\alpha_{\rm wp}$ and $\beta_{\rm wp}$.
The source was assumed to be detected if the TS exceeded 25. 
Then we merged the data to get a densely-sampled composite light curve, shown in Fig.~\ref{fig:gamop}, starting from the 1-h binned light curve and filling the time gaps by adding data from light curves with progressively larger bins, however respecting a time distance between the data points that depends on their light curve bin. 

\begin{figure}
	\includegraphics[width=\columnwidth]{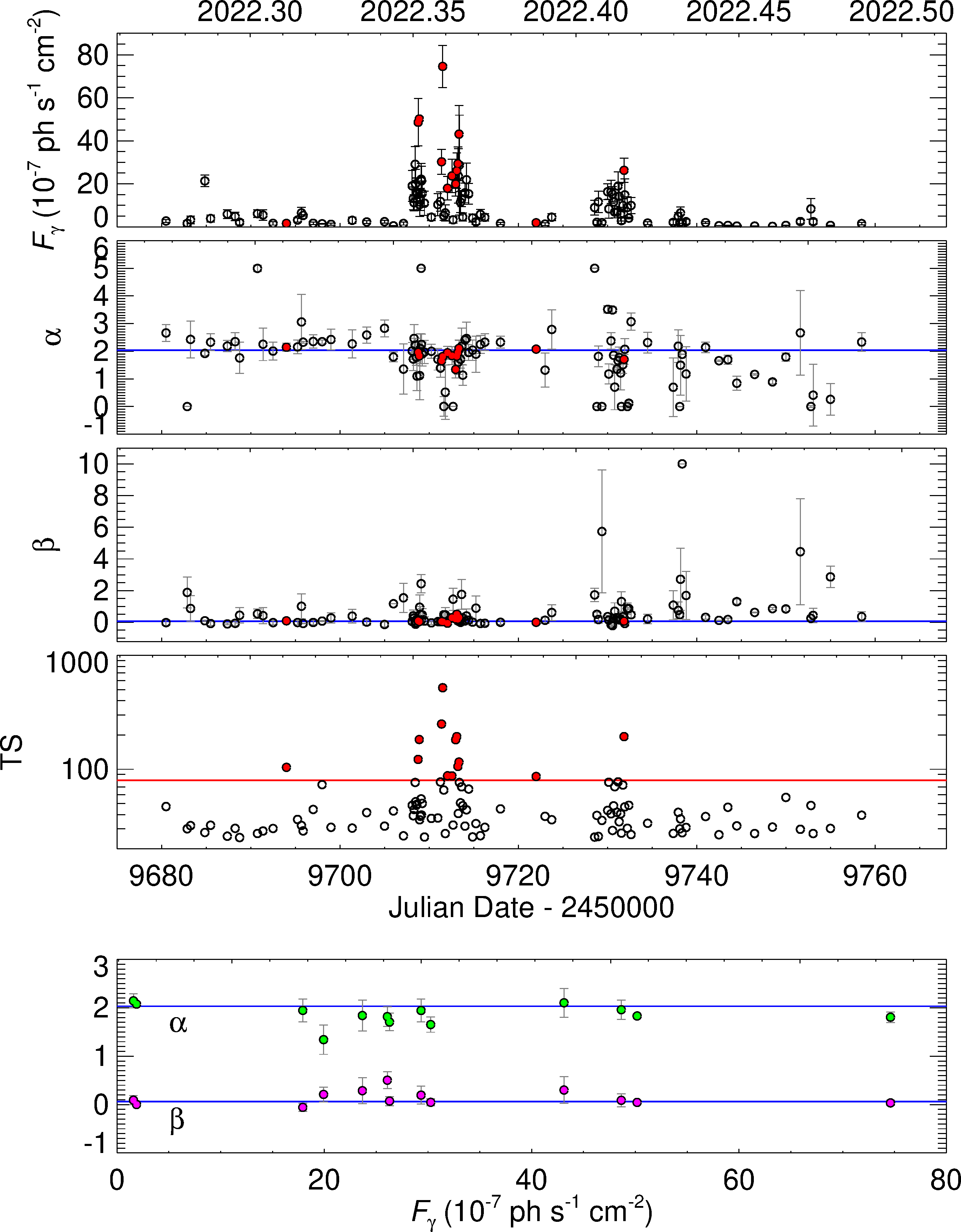}
    \caption{From top to bottom: Composite $\gamma$-ray light curve in the case of variable spectral parameters, the behaviour in time of $\alpha$, $\beta$ and TS, $\alpha$ and $\beta$ versus flux. The blue horizontal lines indicate the values of $\alpha_{\rm wp}$ and $\beta_{\rm wp}$; the red line marks the limit $\rm TS=80$ over which the values of $\alpha$ and $\beta$ are assumed to be reliable (red filled symbols in the $\alpha$ and $\beta$ versus time plots). In the bottom panel, only the $\alpha$ (green) and $\beta$ (magenta) values corresponding to cases with $\rm TS > 80$ are shown.}
    \label{fig:alfa_beta}
\end{figure}

In order to investigate possible spectral changes that are known to occur in blazars, we repeated the whole procedure letting the spectral parameters free to vary. The corresponding composite light curve is shown in Fig.~\ref{fig:gamop} and is very close to the previous one. The behaviour in time of the spectral parameters is displayed in Fig.~\ref{fig:alfa_beta}.
They show a large dispersion, sometimes reaching the boundaries set by the procedure (0 and 5 for $\alpha$, $-5$ and 10 for $\beta$).
We conservatively set $\rm TS=80$ as the limit above which the values of the parameters can be trusted. By plotting these versus flux, we see that they are mostly in agreement with $\alpha_{\rm wp}$ and $\beta_{\rm wp}$, and we cannot recognise any spectral trend.

\subsection{Adaptive binning method}
The adaptive binning $\gamma$-ray light curve was computed using the standard \texttt{Fermitools} software\footnote{\url{https://fermi.gsfc.nasa.gov/ssc/data/analysis/software/}} version 2.2.0 packaged within a \texttt{FermiBottle} container\footnote{\url{https://github.com/fermi-lat/FermiBottle}}. During the analysis we used the same instrument response function, Galactic diffuse emission model, and isotropic background
model as in the fixed binning method described above.
The computations were performed in the unbinned likelihood regime in the 0.1--200 GeV energy range.

The background model includes all the sources from the 4FGL catalogue that fall within $15\degr$ radius around the target location. The fluxes of the background objects within $10\degr$ radius were set as free parameters of the model if their significance is higher than 5.0 according to the 4FGL catalogue. The fluxes of more distant or less significant sources were fixed to their mean values according to the 4FGL catalogue.
The flux of the target itself was modelled using a log-parabolic spectral energy distribution (Eq.~\ref{eq:logparabola}), with the spectral parameters fixed to their catalogue values 
($\alpha_{\rm cat} = 2.125$, $\beta_{\rm cat} = 0.052$, $E_{\rm b} = 674.472 \, \rm MeV$), 
and only the normalisation factor $N_0$ was set as a free parameter to compute the flux.

In order to obtain the highest possible temporal resolution of the light curve, we used an adaptive temporal binning strategy such that the periods of active
state are covered with shorter time bins to obtain a fine structure of the time variability, while quieter states with low flux are covered with wider bins to obtain a
signal-to-noise ratio not lower than some predefined value. In order to do so, we start the integration with a time bin as short as one hour and increase it gradually by 15 min until we reach the test statistic value $\rm TS=25$ (which roughly corresponds to $\sigma=5$). When the desired
TS value is reached, we stop the integration, save the current flux parameters and start a new bin until the whole time range is covered.

The adaptive binning $\gamma$-ray light curve is shown in Fig.~\ref{fig:gamop}; the time bins range from 1 h to about 6 d. The number of epochs is 163, while the composite light curve includes 102 epochs in the case of fixed parameters and 108 in the case of the variable parameters. Therefore, for the following analysis we will use the adaptive binning light curve.

\section{Comparison between the optical and $\gamma$-ray fluxes}
\label{sec:gamo}

The correlation between the blazar flux variations at $\gamma$-ray and optical frequencies has been extensively investigated, especially after the launch of {\it Fermi}, which is scanning the sky every $\sim 3$ hours. The topic is of great interest, since it can provide clues on the origin of the $\gamma$-ray radiation, i.e.\ whether a leptonic or hadronic mechanism is more plausible \citep{dejaeger2023} and, in the first case, what is the nature of the soft seed photons \citep{cohen2014}. However, the results are not always in compliance with the theoretical predictions, and can vary from source to source, and even for the same source observed in different periods \citep{raiteri2012}. In most cases, a strong correlation is found, with (nearly) simultaneous variations in the two bands, as predicted by leptonic models \citep[e.g.][]{raiteri2011,raiteri2013,hovatta2014,larionov2016,carnerero2017,dammando2019}. 
However, in several cases the $\gamma$-ray variations were observed to lead those in the optical band, especially in FSRQs \citep{hayashida2012,cohen2014,carnerero2015,larionov2020}, but there were also cases where the $\gamma$-ray changes appeared delayed \citep{jorstad2011,cohen2014}.

A visual comparison between the optical and $\gamma$-ray fluxes of S4~0954+65 in Fig.~\ref{fig:gamop} shows a general good agreement, with the exception of the first optical flaring phase at $\rm JD \sim 2459680$--2459699, which has only a minor $\gamma$-ray counterpart.

\begin{figure}
\includegraphics[width=\columnwidth]{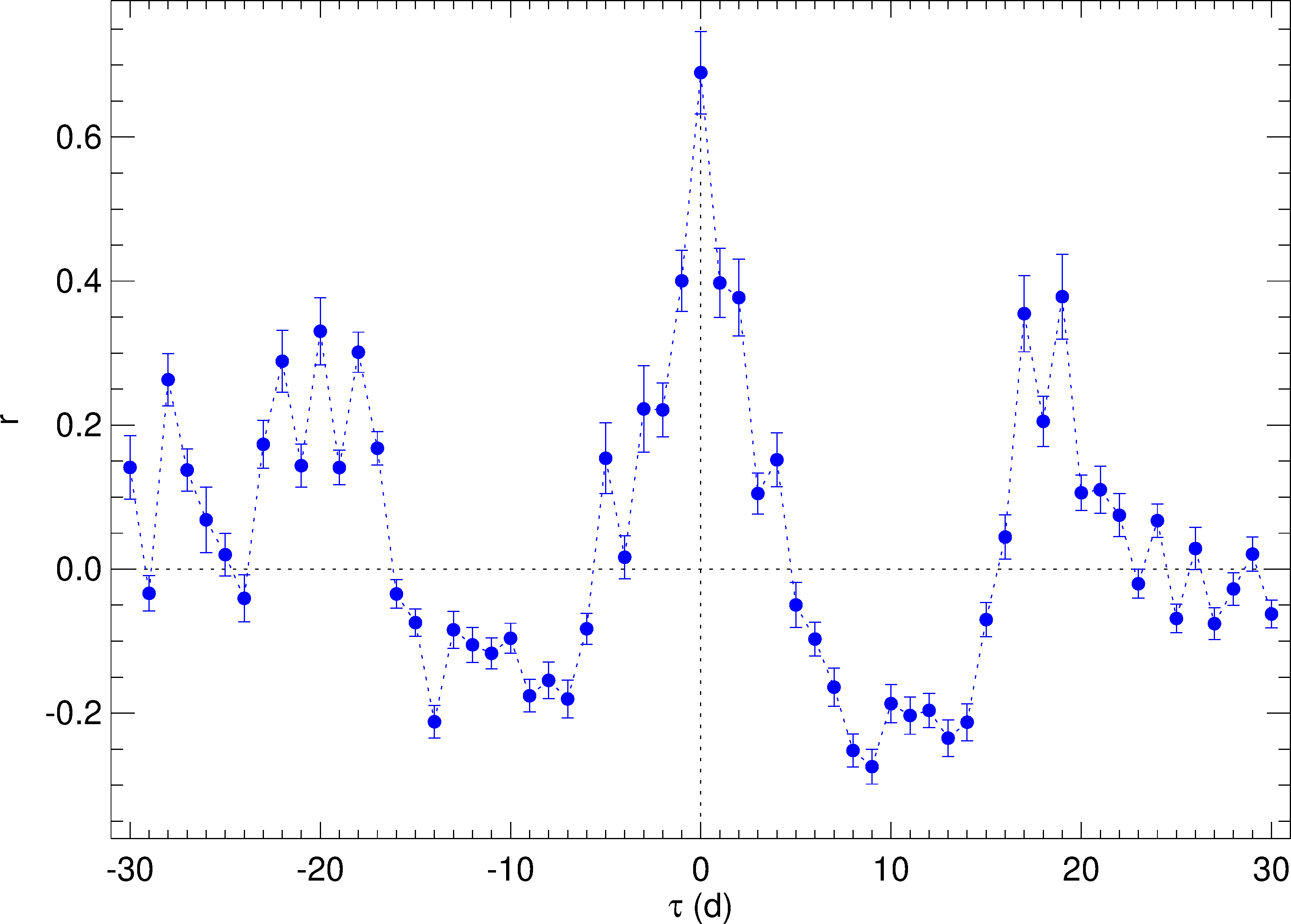}
    \caption{DCF between the adaptive binning $\gamma$-ray light curve and the $R$-band flux densities.}
    \label{fig:dcf_go}
\end{figure}

We investigate the $\gamma$-optical correlation with the DCF.
The result is shown in Fig.~\ref{fig:dcf_go}, where the optical data have previously been averaged over 1 h and the DCF is calculated over 1 d bins. The main peak at $\rm r \sim 0.7$ indicates fair correlation with no time delay (time lag $\tau=0 \rm \, d$).
This suggests that $\gamma$-ray photons can indeed be produced by inverse-Compton scattering of soft photons off the same relativistic electrons that are responsible for the optical synchrotron photons, as predicted by leptonic models.

\begin{figure}
\includegraphics[width=\columnwidth]{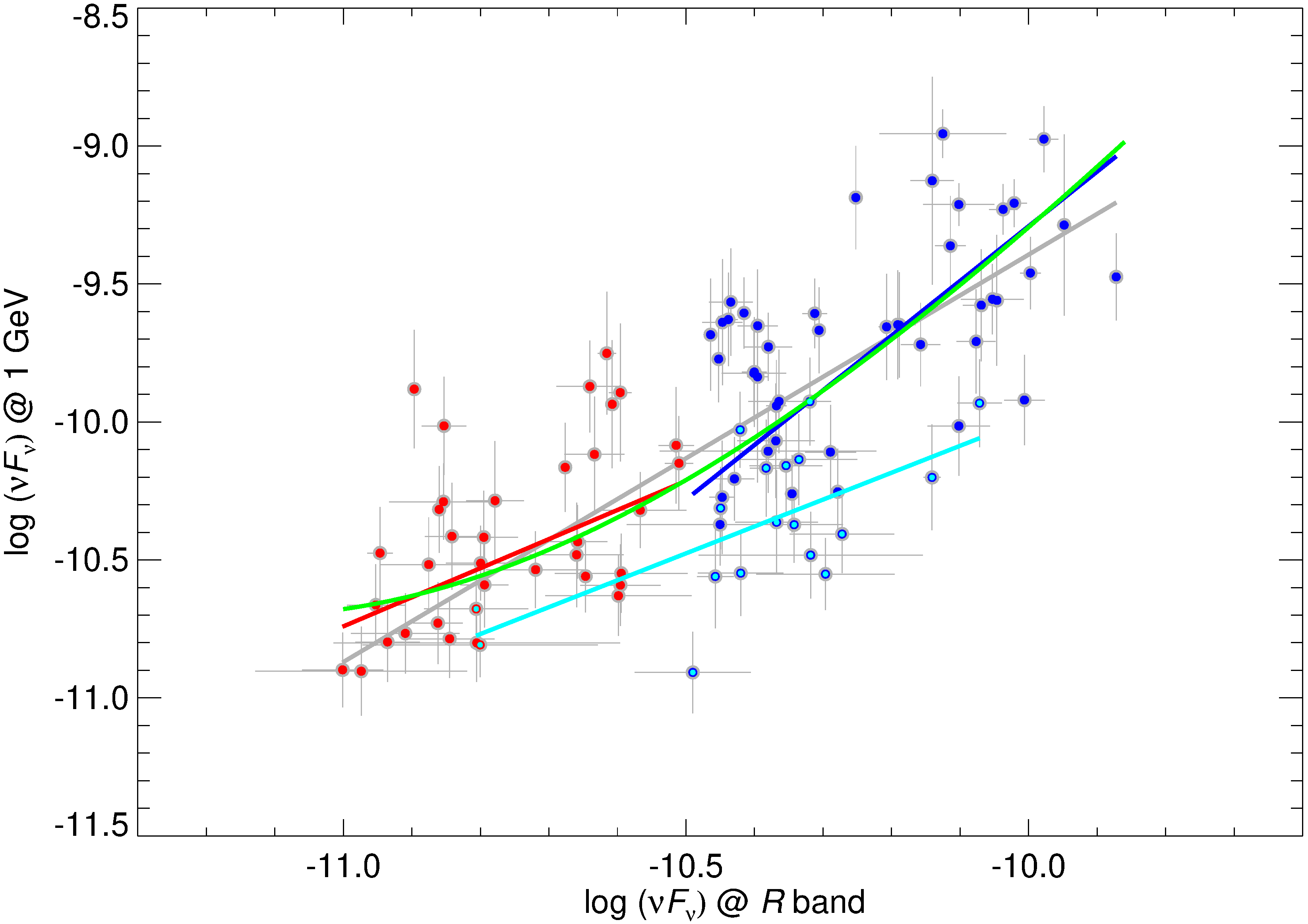}
    \caption{Gamma-ray fluxes at 1 GeV versus the $R$-band fluxes. The red/blue data points correspond to optical data lower/higher than $\log \nu F_\nu = -10.5$. Cyan-filled symbols represent data before $\rm JD=2459699$. The grey, blue, red, and cyan lines are linear regressions to: all the data points, the brightest data points, the faintest data points, and the data points up to $\rm JD=2459699$, respectively. The green line shows a cubic fit to all the data points.
    Vertical grey bars represent errors on the $\gamma$-ray fluxes; horizontal bars indicate standard deviations on the mean of the optical fluxes paired with the $\gamma$-ray ones.}
    \label{fig:gamma-ottico}
\end{figure}

A deeper analysis can reveal further details on the source variability and on the nature of the soft photons seeds.
We transformed the $\gamma$-ray fluxes from counts to physical units taking into account that 
\begin{equation}
\nu F_\nu = 1.602 \times 10^{-3} E^2 {\rm d}N/{\rm d}E \rm \, [erg \, cm^{-2} \, s^{-1}]
\end{equation}
where ${\rm d}N/{\rm d}E$ is given in Eq.~\ref{eq:logparabola}.
The  $\gamma$-ray fluxes at 1 GeV ($\log \nu=23.383$) were then compared to the optical fluxes in the $R$ band ($\log \nu=14.670$). We paired each $\gamma$-ray data point of the adaptive binning light curve with the mean of the optical fluxes included in the same bin interval.
In this way we obtained 97 $\gamma$-optical pairs, where for each $\gamma$-ray data point we averaged from 1 to 470 optical points.
The result is shown in Fig.~\ref{fig:gamma-ottico}. 
The linear regression line on all pairs has a slope of $1.48 \pm 0.05$, but the scatter is rather large. This is actually expected, as two different mechanisms are likely acting. One is the geometrical effect due to the variation of the viewing angle discussed in Section~\ref{sec:doppler}, which would give a slope equal to 1. The other is the synchrotron self-Compton (SSC) process, where the soft photons that are inverse-Compton scattered up to $\gamma$-ray energies are the same optical photons \citep{maraschi1992,bloom1996}. This would yield a slope of 2. Because of the presence of these two mechanisms, the distribution of data points would approximately lie within a parallelogram, as shown by  \citet{larionov2016}. Therefore, a cubic regression (see Fig.~\ref{fig:gamma-ottico}) is a more suitable fit, since it can follow the different slopes at the various brightness states.

To check for the double nature of the $\gamma$-optical correlation, we considered high and low brightness states separately. The slope of the linear regression on the data points corresponding to the brightest optical states only ($\log \nu F_\nu > -10.5$, see Fig.~\ref{fig:gamop}) is $1.98 \pm 0.12$, while in the case of the faintest states the slope decreases to $1.05 \pm 0.19$.
This matches well the trend of the cubic fit and can be understood as follows.
In faint states, longer integration intervals are required to get significant $\gamma$-ray signals, and also the optical data to be paired with them are consequently averaged on long time bins. 
This is highlighted by the large horizontal bars in Fig.~\ref{fig:gamma-ottico}, which represent the standard deviations of the optical data averaged in the time bin of the corresponding $\gamma$-ray data point. Therefore, the fast, intrinsic flux changes are smoothed out, and the slope becomes close to 1, in agreement with a geometrical origin of the long-term flux variations \citep{raiteri2017_nature}. We note that a similar slope is found during the first optical active phase, which has only a minor $\gamma$-ray counterpart, which smooths the variability. Indeed, the optical-$\gamma$ correlation before $\rm JD=2459699$ has a power-law index $0.97 \pm 0.18$ (see Fig.~\ref{fig:gamma-ottico}).
In contrast, during the brightest phases, the time bins become shorter and it is possible to appreciate the squared dependence of the $\gamma$-ray fluxes on the optical ones due to the SSC process, whose signature prevails over the geometric effect.

\section{Discussion and conclusions}
We have presented the photometric and polarimetric optical data obtained by the WEBT Collaboration during a very active phase of the BL Lac-type object S4~0954+65 in 2022, together with the $\gamma$-ray data from the {\it Fermi} satellite.
In this period the source reached its historical brightness maxima in both optical and $\gamma$-ray band.

Many unprecedented, extreme episodes of optical intraday variability were detected. A model fit to one of the fastest flares implies a variability time scale of about 17 min and thus a size of the optical emitting region less than $3 \times 10^{14} \, \rm cm$ (about $10^{-4}$ parsec). This means that we are likely observing emission coming from jet sub-regions, confirming earlier suggestions arising from the detection of minute-scale variability also at GeV \citep{shukla2020} and TeV energies \citep{aharonian2007,albert2007}.

We have analysed the behaviour of the optical polarization, which shows large fluctuations in both the polarization degree $P$ and angle, and a general anticorrelation between $P$ and the optical flux density $F_R$. 
The presence of this defined trend together with strong dispersion around it may indicate that we are seeing the combination of different processes.

We have shown that if the long-term trend of the optical flux density  is due to a variation of the Doppler factor $\delta$ caused by orientation changes of the jet emitting region, then a simple helical magnetic field in a possibly rotating jet can well describe the observed average behaviour of $P$. The same result can also be obtained in the framework of a shock-in-jet model.
All the polarization models that we considered predict that the trend of $P$ follows that of the viewing angle $\theta$. Therefore, since $\theta$ anticorrelates with $\delta$, which in our view determines the long-term behaviour of the source brightness, these models naturally explain the general anticorrelation between $P$ and $F_R$.

However, there are periods in which the observed $P$ is lower than the average trend predicted by the models, and periods where it is higher. 
One way to explain the whole range of $P$ values is to assume a variation of the model parameters, such as the jet angular rotational velocity or the strength of the shock waves.
More realistically, when $P$ becomes much lower than predicted, 
we can imagine that some process is leading to a less ordered magnetic field, most likely increasing its turbulent component. 
This would also be consistent with our finding that the emission comes from jet sub-regions, because turbulence can be seen as the overlapping of multiple stochastic contributions \citep[e.g.][]{marscher2014}.
In contrast, the observation of values of $P$ higher than predicted suggests that something has produced a more ordered field. 
This can occur when shock waves propagate in the jet, compressing the magnetic field lines \citep{hughes1985,marscher1985}, or in any other case when the field lines appear more parallel along the line of sight through the jet.

An alternative mechanism that can produce flux and polarization variability is magnetic reconnection. According to the simulations by \citet{zhang2020}, who explored the magnetic reconnection flux and polarization signatures in relativistic jets, the outcome of magnetic reconnection strongly depends on the model parameters, i.e.\ on the physical conditions. In particular, the value of the polarization degree is very sensitive to the strength of the guide field, which represents the component of the magnetic field which is perpendicular to the reconnecting magnetic field. 
This mechanism could also explain fast flux variability from small size emitting regions, since reconnection can lead to the formation of a large number of plasmoids \citep{giannios2013}.

The comparison between optical and $\gamma$-ray data leads to results that are consistent with our geometrical interpretation of the long-term trend. The optical and $\gamma$-ray flux variations are well correlated with no appreciable time delay, which means that the observed optical and $\gamma$-ray photons come from the same jet region. Moreover, the power-law dependence of the $\gamma$-ray fluxes on the optical ones has an index $\approx 1$ during faint states, where the variability is dominated by the long-term trend because of the low $\gamma$-ray statistics. An index one is what is expected if the long-term trend is due to orientation changes. In contrast, the index becomes $\approx 2$ in bright states, when the higher statistics allow us to detect what is most likely the signature of the intrinsic SSC process.
We note that an SSC nature of the $\gamma$-ray emission in S4~0954+65 is in agreement with previous results that favour an SSC mechanism for BL Lacs in general, and explains the $\gamma$-optical correlation with no appreciable time delay \citep[e.g.,][and references therein]{cohen2014,hovatta2014}.

In conclusion, we interpret the long-term optical flux and polarization behaviour as the result of variations in the viewing angle of the optical emitting region.
The short-term variability would instead be produced by energetic processes occurring inside the jet. Polarimetry suggests that there are periods where turbulence dominates the optical emitting region, while in other periods the magnetic field becomes more ordered maybe because of the passage of a shock wave. Magnetic reconnection could also be a viable explanation for the short-term photometric and polarimetric variability.
The $\gamma$-ray emission, which comes from the same emitting region, seems to confirm the geometric nature of the long-term trend, and during the brightest states reveals its SSC nature.

\section*{Acknowledgements}
We are indebted to Leslie Brown, Dmitry Blinov, Eduard V. Emelianov, Alexander S.Moskvitin, Timur A. Fatkhullin, Olga A. Maslennikova, Alexis Foretier, Lorenzo Pizzuti.
This research has made use of NASA’s Astrophysics Data System Bibliographic Services.
HG and KW acknowledge financial support from the Department of Science and Technology, India, through an INSPIRE Faculty Award IFA17-PH197 at ARIES, Nainital, India.
The $R$-band photometric data from the University of Athens Observatory 
(UOAO) were obtained after utilizing the robotic and remotely 
controlled instruments at the facilities \citep{gazeas2016}.
This research was partially supported by the Bulgarian National Science Fund of the Ministry of Education and Science under grants KP-06-H38/4 (2019), KP-06-KITAJ/2 (2020) and KP-06-H68/4 (2022). 
KM acknowledges support from JSPS KAKENHI grant number 19K03930.
This work is supported by the Ministry of Science and Higher Education of the Russian Federation under the contract 075-15-2022-262. The SAO-RAS observations were carried out with the Big Telescope Alt-azimuthal facility.
S.K. acknowledges support from the European Research Council (ERC) under the European Union Horizon 2020 research and innovation program under the grant agreement No 771282.
GD, OV, MDJ and MS acknowledge support by the Astronomical station Vidojevica, funding from the Ministry of Science, Technological Development and Innovation of the Republic of Serbia
(contract No. 451-03-47/2023-01/200002), by the EC through project
BELISSIMA (call FP7-REGPOT-2010-5, No. 265772), the observing and
financial grant support from the Institute of Astronomy and Rozhen NAO
BAS through the bilateral SANU-BAN joint research project ``GAIA
astrometry and fast variable astronomical objects", and support by the
SANU project F-187.
MDJ thanks the Brigham Young University Department of Physics and Astronomy for continued support of the ongoing AGN monitoring program at the West Mountain Observatory.
JAP acknowledge financial support from the Spanish Ministry of Science and Innovation (MICINN) through the Spanish State Research Agency, under Severo Ochoa Program 2020- 2023 (CEX2019-000920-S).
This paper is partly based on observations made with the IAC-80 telescope operated on the island of Tenerife by the Instituto de Astrof'isica de Canarias in the Spanish Observatorio del Teide and on observations made with the LCOGT 0.4 m telescope network, one of whose nodes is located in the Spanish Observatorio del Teide.
EB acknowledges support from DGAPA-PAPIIT GRANT IN119123.
This work is partly based upon observations carried out at the Observatorio Astron\'omico Nacional 
on the Sierra San Pedro M\'artir (OAN-SPM), Baja California, Mexico.
VJ acknowledges the support provided by the Department of Science and Technology (DST) under the ``Fund for Improvement of S \& T Infrastructure (FIST)" program (SR/FST/PS-I/2022/208).

\section*{Data Availability}

    Data acquired by the WEBT collaboration are stored in the WEBT archive and are available upon request to the WEBT President Massimo Villata (\href{mailto:massimo.villata@inaf.it}{massimo.villata@inaf.it}). Fermi data can be downloaded from the National Aeronautics and Space Administration (NASA) site (\url{https://fermi.gsfc.nasa.gov/ssc/data/access/}) and analysed with the Fermitools software ( \url{https://fermi.gsfc.nasa.gov/ssc/data/analysis/documentation/}); the $\gamma$-ray light curves published here can be obtained from the authors under request.



\bibliographystyle{mnras}
\bibliography{flare} 

\begin{thebibliography}{}
\makeatletter
\relax
\def\mn@urlcharsother{\let\do\@makeother \do\$\do\&\do\#\do\^\do\_\do\%\do\~}
\def\mn@doi{\begingroup\mn@urlcharsother \@ifnextchar [ {\mn@doi@}
  {\mn@doi@[]}}
\def\mn@doi@[#1]#2{\def\@tempa{#1}\ifx\@tempa\@empty \href
  {http://dx.doi.org/#2} {doi:#2}\else \href {http://dx.doi.org/#2} {#1}\fi
  \endgroup}
\def\mn@eprint#1#2{\mn@eprint@#1:#2::\@nil}
\def\mn@eprint@arXiv#1{\href {http://arxiv.org/abs/#1} {{\tt arXiv:#1}}}
\def\mn@eprint@dblp#1{\href {http://dblp.uni-trier.de/rec/bibtex/#1.xml}
  {dblp:#1}}
\def\mn@eprint@#1:#2:#3:#4\@nil{\def\@tempa {#1}\def\@tempb {#2}\def\@tempc
  {#3}\ifx \@tempc \@empty \let \@tempc \@tempb \let \@tempb \@tempa \fi \ifx
  \@tempb \@empty \def\@tempb {arXiv}\fi \@ifundefined
  {mn@eprint@\@tempb}{\@tempb:\@tempc}{\expandafter \expandafter \csname
  mn@eprint@\@tempb\endcsname \expandafter{\@tempc}}}

\bibitem[\protect\citeauthoryear{{Abdollahi} et~al.,}{{Abdollahi}
  et~al.}{2023}]{abdollahi2023}
{Abdollahi} S.,  et~al., 2023, \mn@doi [\apjs] {10.3847/1538-4365/acbb6a},
  \href {https://ui.adsabs.harvard.edu/abs/2023ApJS..265...31A} {265, 31}

\bibitem[\protect\citeauthoryear{{Aharonian} et~al.,}{{Aharonian}
  et~al.}{2007}]{aharonian2007}
{Aharonian} F.,  et~al., 2007, \mn@doi [\apjl] {10.1086/520635}, \href
  {https://ui.adsabs.harvard.edu/abs/2007ApJ...664L..71A} {664, L71}

\bibitem[\protect\citeauthoryear{{Albert} et~al.,}{{Albert}
  et~al.}{2007}]{albert2007}
{Albert} J.,  et~al., 2007, \mn@doi [\apj] {10.1086/521382}, \href
  {https://ui.adsabs.harvard.edu/abs/2007ApJ...669..862A} {669, 862}

\bibitem[\protect\citeauthoryear{{Atwood} et~al.,}{{Atwood}
  et~al.}{2009}]{atwood2009}
{Atwood} W.~B.,  et~al., 2009, \mn@doi [\apj] {10.1088/0004-637X/697/2/1071},
  \href {https://ui.adsabs.harvard.edu/abs/2009ApJ...697.1071A} {697, 1071}

\bibitem[\protect\citeauthoryear{{Bachev}}{{Bachev}}{2015}]{bachev2015}
{Bachev} R.,  2015, \mn@doi [\mnras] {10.1093/mnrasl/slv059}, \href
  {https://ui.adsabs.harvard.edu/abs/2015MNRAS.451L..21B} {451, L21}

\bibitem[\protect\citeauthoryear{{Bachev} \& {Strigachev}}{{Bachev} \&
  {Strigachev}}{2022}]{bachev2022}
{Bachev} R.,  {Strigachev} A.,  2022, The Astronomer's Telegram, \href
  {https://ui.adsabs.harvard.edu/abs/2022ATel15322....1B} {15322, 1}

\bibitem[\protect\citeauthoryear{{Becerra Gonz{\'a}lez}, {Acosta-Pulido},
  {Boschin}, {Clavero}, {Otero-Santos}, {Carballo-Bello}  \&
  {Dom{\'\i}nguez-Palmero}}{{Becerra Gonz{\'a}lez} et~al.}{2021}]{becerra2021}
{Becerra Gonz{\'a}lez} J.,  {Acosta-Pulido} J.~A.,  {Boschin} W.,  {Clavero}
  R.,  {Otero-Santos} J.,  {Carballo-Bello} J.~A.,   {Dom{\'\i}nguez-Palmero}
  L.,  2021, \mn@doi [\mnras] {10.1093/mnras/stab1274}, \href
  {https://ui.adsabs.harvard.edu/abs/2021MNRAS.504.5258B} {504, 5258}

\bibitem[\protect\citeauthoryear{{Bessell}, {Castelli}  \& {Plez}}{{Bessell}
  et~al.}{1998}]{bessell1998}
{Bessell} M.~S.,  {Castelli} F.,   {Plez} B.,  1998, \aap, \href
  {http://adsabs.harvard.edu/cgi-bin/nph-bib_query?bibcode=1998A%26A...333..231B&db_key=AST}
  {333, 231}

\bibitem[\protect\citeauthoryear{{Bloom} \& {Marscher}}{{Bloom} \&
  {Marscher}}{1996}]{bloom1996}
{Bloom} S.~D.,  {Marscher} A.~P.,  1996, \mn@doi [\apj] {10.1086/177092}, \href
  {https://ui.adsabs.harvard.edu/abs/1996ApJ...461..657B} {461, 657}

\bibitem[\protect\citeauthoryear{{Bodo}, {Tavecchio}  \& {Sironi}}{{Bodo}
  et~al.}{2021}]{bodo2021}
{Bodo} G.,  {Tavecchio} F.,   {Sironi} L.,  2021, \mn@doi [\mnras]
  {10.1093/mnras/staa3620}, \href
  {https://ui.adsabs.harvard.edu/abs/2021MNRAS.501.2836B} {501, 2836}

\bibitem[\protect\citeauthoryear{{B{\"o}ttcher}, {Reimer}, {Sweeney}  \&
  {Prakash}}{{B{\"o}ttcher} et~al.}{2013}]{boettcher2013}
{B{\"o}ttcher} M.,  {Reimer} A.,  {Sweeney} K.,   {Prakash} A.,  2013, \mn@doi
  [\apj] {10.1088/0004-637X/768/1/54}, \href
  {http://esoads.eso.org/abs/2013ApJ...768...54B} {768, 54}

\bibitem[\protect\citeauthoryear{{Carnerero} et~al.,}{{Carnerero}
  et~al.}{2015}]{carnerero2015}
{Carnerero} M.~I.,  et~al., 2015, \mn@doi [\mnras] {10.1093/mnras/stv823},
  \href {https://ui.adsabs.harvard.edu/abs/2015MNRAS.450.2677C} {450, 2677}

\bibitem[\protect\citeauthoryear{{Carnerero} et~al.,}{{Carnerero}
  et~al.}{2017}]{carnerero2017}
{Carnerero} M.~I.,  et~al., 2017, \mn@doi [\mnras] {10.1093/mnras/stx2185},
  \href {https://ui.adsabs.harvard.edu/abs/2017MNRAS.472.3789C} {472, 3789}

\bibitem[\protect\citeauthoryear{{Cohen}, {Romani}, {Filippenko}, {Cenko},
  {Lott}, {Zheng}  \& {Li}}{{Cohen} et~al.}{2014}]{cohen2014}
{Cohen} D.~P.,  {Romani} R.~W.,  {Filippenko} A.~V.,  {Cenko} S.~B.,  {Lott}
  B.,  {Zheng} W.,   {Li} W.,  2014, \mn@doi [\apj]
  {10.1088/0004-637X/797/2/137}, \href
  {https://ui.adsabs.harvard.edu/abs/2014ApJ...797..137C} {797, 137}

\bibitem[\protect\citeauthoryear{{D'Ammando} et~al.,}{{D'Ammando}
  et~al.}{2019}]{dammando2019}
{D'Ammando} F.,  et~al., 2019, \mn@doi [\mnras] {10.1093/mnras/stz2792}, \href
  {https://ui.adsabs.harvard.edu/abs/2019MNRAS.490.5300D} {490, 5300}

\bibitem[\protect\citeauthoryear{{Edelson} \& {Krolik}}{{Edelson} \&
  {Krolik}}{1988}]{edelson1988}
{Edelson} R.~A.,  {Krolik} J.~H.,  1988, \mn@doi [\apj] {10.1086/166773}, \href
  {http://adsabs.harvard.edu/cgi-bin/nph-bib_query?bibcode=1988ApJ...333..646E&db_key=AST}
  {333, 646}

\bibitem[\protect\citeauthoryear{{Gabuzda}, {Kochenov}, {Kollgaard}  \&
  {Cawthorne}}{{Gabuzda} et~al.}{2000}]{gabuzda2000}
{Gabuzda} D.~C.,  {Kochenov} P.~Y.,  {Kollgaard} R.~I.,   {Cawthorne} T.~V.,
  2000, \mn@doi [\mnras] {10.1046/j.1365-8711.2000.03400.x}, \href
  {https://ui.adsabs.harvard.edu/abs/2000MNRAS.315..229G} {315, 229}

\bibitem[\protect\citeauthoryear{{Gazeas}}{{Gazeas}}{2016}]{gazeas2016}
{Gazeas} K.,  2016, in Revista Mexicana de Astronomia y Astrofisica Conference
  Series. pp 22--23

\bibitem[\protect\citeauthoryear{{Ghisellini}, {Celotti}, {Fossati}, {Maraschi}
   \& {Comastri}}{{Ghisellini} et~al.}{1998}]{ghisellini1998}
{Ghisellini} G.,  {Celotti} A.,  {Fossati} G.,  {Maraschi} L.,   {Comastri} A.,
   1998, \mn@doi [\mnras] {10.1046/j.1365-8711.1998.02032.x}, \href
  {https://ui.adsabs.harvard.edu/abs/1998MNRAS.301..451G} {301, 451}

\bibitem[\protect\citeauthoryear{{Giannios}}{{Giannios}}{2013}]{giannios2013}
{Giannios} D.,  2013, \mn@doi [\mnras] {10.1093/mnras/stt167}, \href
  {https://ui.adsabs.harvard.edu/abs/2013MNRAS.431..355G} {431, 355}

\bibitem[\protect\citeauthoryear{{Gong}, {Tian}, {Zhou}, {Yi}  \&
  {Fang}}{{Gong} et~al.}{2023}]{gong2023}
{Gong} Y.,  {Tian} S.,  {Zhou} L.,  {Yi} T.,   {Fang} J.,  2023, \mn@doi [\apj]
  {10.3847/1538-4357/acca7b}, \href
  {https://ui.adsabs.harvard.edu/abs/2023ApJ...949...39G} {949, 39}

\bibitem[\protect\citeauthoryear{{Hagen-Thorn} et~al.,}{{Hagen-Thorn}
  et~al.}{2015}]{hagen2015}
{Hagen-Thorn} V.~A.,  et~al., 2015, \mn@doi [Astronomy Reports]
  {10.1134/S1063772915050030}, \href
  {https://ui.adsabs.harvard.edu/abs/2015ARep...59..551H} {59, 551}

\bibitem[\protect\citeauthoryear{{Hayashida} et~al.,}{{Hayashida}
  et~al.}{2012}]{hayashida2012}
{Hayashida} M.,  et~al., 2012, \mn@doi [\apj] {10.1088/0004-637X/754/2/114},
  \href {https://ui.adsabs.harvard.edu/abs/2012ApJ...754..114H} {754, 114}

\bibitem[\protect\citeauthoryear{{Heidt} \& {Wagner}}{{Heidt} \&
  {Wagner}}{1996}]{heidt1996}
{Heidt} J.,  {Wagner} S.~J.,  1996, \mn@doi [\aap]
  {10.48550/arXiv.astro-ph/9506032}, \href
  {https://ui.adsabs.harvard.edu/abs/1996A&A...305...42H} {305, 42}

\bibitem[\protect\citeauthoryear{{Hovatta} et~al.,}{{Hovatta}
  et~al.}{2014}]{hovatta2014}
{Hovatta} T.,  et~al., 2014, \mn@doi [\mnras] {10.1093/mnras/stt2494}, \href
  {https://ui.adsabs.harvard.edu/abs/2014MNRAS.439..690H} {439, 690}

\bibitem[\protect\citeauthoryear{{Hufnagel} \& {Bregman}}{{Hufnagel} \&
  {Bregman}}{1992}]{hufnagel1992}
{Hufnagel} B.~R.,  {Bregman} J.~N.,  1992, \mn@doi [\apj] {10.1086/171033},
  \href
  {http://adsabs.harvard.edu/cgi-bin/nph-bib_query?bibcode=1992ApJ...386..473H&db_key=AST}
  {386, 473}

\bibitem[\protect\citeauthoryear{{Hughes}, {Aller}  \& {Aller}}{{Hughes}
  et~al.}{1985}]{hughes1985}
{Hughes} P.~A.,  {Aller} H.~D.,   {Aller} M.~F.,  1985, \mn@doi [\apj]
  {10.1086/163611}, \href
  {https://ui.adsabs.harvard.edu/abs/1985ApJ...298..301H} {298, 301}

\bibitem[\protect\citeauthoryear{{Jorstad}, {Marscher}, {Agudo}, {Smith},
  {Larionov}  \& {L{\"a}hteenm{\"a}ki}}{{Jorstad} et~al.}{2011}]{jorstad2011}
{Jorstad} S.~G.,  {Marscher} A.~P.,  {Agudo} I.,  {Smith} P.~S.,  {Larionov}
  V.~M.,   {L{\"a}hteenm{\"a}ki} A.,  2011, \mn@doi [Journal of Astrophysics
  and Astronomy] {10.1007/s12036-011-9038-z}, \href
  {https://ui.adsabs.harvard.edu/abs/2011JApA...32..239J} {32, 239}

\bibitem[\protect\citeauthoryear{{Jorstad} et~al.,}{{Jorstad}
  et~al.}{2017}]{jorstad2017}
{Jorstad} S.~G.,  et~al., 2017, \mn@doi [\apj] {10.3847/1538-4357/aa8407},
  \href {https://ui.adsabs.harvard.edu/abs/2017ApJ...846...98J} {846, 98}

\bibitem[\protect\citeauthoryear{{Jorstad} et~al.,}{{Jorstad}
  et~al.}{2022}]{jorstad2022}
{Jorstad} S.~G.,  et~al., 2022, \mn@doi [\nat] {10.1038/s41586-022-05038-9},
  \href {https://ui.adsabs.harvard.edu/abs/2022Natur.609..265J} {609, 265}

\bibitem[\protect\citeauthoryear{{Kishore}, {Gupta}  \& {Wiita}}{{Kishore}
  et~al.}{2023}]{kishore2023}
{Kishore} S.,  {Gupta} A.~C.,   {Wiita} P.~J.,  2023, \mn@doi [\apj]
  {10.3847/1538-4357/aca809}, \href
  {https://ui.adsabs.harvard.edu/abs/2023ApJ...943...53K} {943, 53}

\bibitem[\protect\citeauthoryear{{Larionov} et~al.,}{{Larionov}
  et~al.}{2013}]{larionov2013}
{Larionov} V.~M.,  et~al., 2013, \mn@doi [\apj] {10.1088/0004-637X/768/1/40},
  \href {https://ui.adsabs.harvard.edu/abs/2013ApJ...768...40L} {768, 40}

\bibitem[\protect\citeauthoryear{{Larionov} et~al.,}{{Larionov}
  et~al.}{2016}]{larionov2016}
{Larionov} V.~M.,  et~al., 2016, \mn@doi [\mnras] {10.1093/mnras/stw1516},
  \href {https://ui.adsabs.harvard.edu/abs/2016MNRAS.461.3047L} {461, 3047}

\bibitem[\protect\citeauthoryear{{Larionov} et~al.,}{{Larionov}
  et~al.}{2020}]{larionov2020}
{Larionov} V.~M.,  et~al., 2020, \mn@doi [\mnras] {10.1093/mnras/staa082},
  \href {https://ui.adsabs.harvard.edu/abs/2020MNRAS.492.3829L} {492, 3829}

\bibitem[\protect\citeauthoryear{{Lyutikov}, {Pariev}  \& {Gabuzda}}{{Lyutikov}
  et~al.}{2005}]{lyutikov2005}
{Lyutikov} M.,  {Pariev} V.~I.,   {Gabuzda} D.~C.,  2005, \mn@doi [\mnras]
  {10.1111/j.1365-2966.2005.08954.x}, \href
  {https://ui.adsabs.harvard.edu/abs/2005MNRAS.360..869L} {360, 869}

\bibitem[\protect\citeauthoryear{{MAGIC Collaboration} et~al.,}{{MAGIC
  Collaboration} et~al.}{2018}]{magic2018}
{MAGIC Collaboration} et~al., 2018, \mn@doi [\aap]
  {10.1051/0004-6361/201832624}, \href
  {https://ui.adsabs.harvard.edu/abs/2018A&A...617A..30M} {617, A30}

\bibitem[\protect\citeauthoryear{{Maraschi}, {Ghisellini}  \&
  {Celotti}}{{Maraschi} et~al.}{1992}]{maraschi1992}
{Maraschi} L.,  {Ghisellini} G.,   {Celotti} A.,  1992, \mn@doi [\apjl]
  {10.1086/186531}, \href
  {https://ui.adsabs.harvard.edu/abs/1992ApJ...397L...5M} {397, L5}

\bibitem[\protect\citeauthoryear{{Marchili}, {Krichbaum}, {Liu}, {Song},
  {Gab{\'a}nyi}, {Fuhrmann}, {Witzel}  \& {Zensus}}{{Marchili}
  et~al.}{2012}]{marchili2012}
{Marchili} N.,  {Krichbaum} T.~P.,  {Liu} X.,  {Song} H.~G.,  {Gab{\'a}nyi}
  K.~{\'E}.,  {Fuhrmann} L.,  {Witzel} A.,   {Zensus} J.~A.,  2012, \mn@doi
  [\aap] {10.1051/0004-6361/201218977}, \href
  {https://ui.adsabs.harvard.edu/abs/2012A&A...542A.121M} {542, A121}

\bibitem[\protect\citeauthoryear{{Marchini} et~al.,}{{Marchini}
  et~al.}{2022}]{marchini2022}
{Marchini} A.,  et~al., 2022, The Astronomer's Telegram, \href
  {https://ui.adsabs.harvard.edu/abs/2022ATel15380....1M} {15380, 1}

\bibitem[\protect\citeauthoryear{{Marscher}}{{Marscher}}{2014}]{marscher2014}
{Marscher} A.~P.,  2014, \mn@doi [\apj] {10.1088/0004-637X/780/1/87}, \href
  {https://ui.adsabs.harvard.edu/abs/2014ApJ...780...87M} {780, 87}

\bibitem[\protect\citeauthoryear{{Marscher} \& {Gear}}{{Marscher} \&
  {Gear}}{1985}]{marscher1985}
{Marscher} A.~P.,  {Gear} W.~K.,  1985, \mn@doi [\apj] {10.1086/163592}, \href
  {https://ui.adsabs.harvard.edu/abs/1985ApJ...298..114M} {298, 114}

\bibitem[\protect\citeauthoryear{{Morozova} et~al.,}{{Morozova}
  et~al.}{2014}]{morozova2014}
{Morozova} D.~A.,  et~al., 2014, \mn@doi [\aj] {10.1088/0004-6256/148/3/42},
  \href {https://ui.adsabs.harvard.edu/abs/2014AJ....148...42M} {148, 42}

\bibitem[\protect\citeauthoryear{{Mukherjee} et~al.,}{{Mukherjee}
  et~al.}{1995}]{mukherjee1995}
{Mukherjee} R.,  et~al., 1995, \mn@doi [\apj] {10.1086/175685}, \href
  {https://ui.adsabs.harvard.edu/abs/1995ApJ...445..189M} {445, 189}

\bibitem[\protect\citeauthoryear{{Papadakis}, {Samaritakis}, {Boumis}  \&
  {Papamastorakis}}{{Papadakis} et~al.}{2004}]{papadakis2004}
{Papadakis} I.~E.,  {Samaritakis} V.,  {Boumis} P.,   {Papamastorakis} J.,
  2004, \mn@doi [\aap] {10.1051/0004-6361:20040446}, \href
  {https://ui.adsabs.harvard.edu/abs/2004A&A...426..437P} {426, 437}

\bibitem[\protect\citeauthoryear{{Pariev}, {Istomin}  \& {Beresnyak}}{{Pariev}
  et~al.}{2003}]{pariev2003}
{Pariev} V.~I.,  {Istomin} Y.~N.,   {Beresnyak} A.~R.,  2003, \mn@doi [\aap]
  {10.1051/0004-6361:20030350}, \href
  {https://ui.adsabs.harvard.edu/abs/2003A&A...403..805P} {403, 805}

\bibitem[\protect\citeauthoryear{{Raiteri} \& {Villata}}{{Raiteri} \&
  {Villata}}{2021}]{raiteri2021_galaxies}
{Raiteri} C.~M.,  {Villata} M.,  2021, \mn@doi [Galaxies]
  {10.3390/galaxies9020042}, \href
  {https://ui.adsabs.harvard.edu/abs/2021Galax...9...42R} {9, 42}

\bibitem[\protect\citeauthoryear{{Raiteri} et~al.,}{{Raiteri}
  et~al.}{1999}]{raiteri1999}
{Raiteri} C.~M.,  et~al., 1999, \aap, \href
  {https://ui.adsabs.harvard.edu/abs/1999A&A...352...19R} {352, 19}

\bibitem[\protect\citeauthoryear{{Raiteri} et~al.,}{{Raiteri}
  et~al.}{2011}]{raiteri2011}
{Raiteri} C.~M.,  et~al., 2011, \mn@doi [\aap] {10.1051/0004-6361/201117026},
  \href {https://ui.adsabs.harvard.edu/abs/2011A&A...534A..87R} {534, A87}

\bibitem[\protect\citeauthoryear{{Raiteri} et~al.,}{{Raiteri}
  et~al.}{2012}]{raiteri2012}
{Raiteri} C.~M.,  et~al., 2012, \mn@doi [\aap] {10.1051/0004-6361/201219492},
  \href {https://ui.adsabs.harvard.edu/abs/2012A&A...545A..48R} {545, A48}

\bibitem[\protect\citeauthoryear{{Raiteri} et~al.,}{{Raiteri}
  et~al.}{2013}]{raiteri2013}
{Raiteri} C.~M.,  et~al., 2013, \mn@doi [\mnras] {10.1093/mnras/stt1672}, \href
  {https://ui.adsabs.harvard.edu/abs/2013MNRAS.436.1530R} {436, 1530}

\bibitem[\protect\citeauthoryear{{Raiteri} et~al.,}{{Raiteri}
  et~al.}{2017}]{raiteri2017_nature}
{Raiteri} C.~M.,  et~al., 2017, \mn@doi [\nat] {10.1038/nature24623}, \href
  {https://ui.adsabs.harvard.edu/abs/2017Natur.552..374R} {552, 374}

\bibitem[\protect\citeauthoryear{{Raiteri} et~al.,}{{Raiteri}
  et~al.}{2021}]{raiteri2021}
{Raiteri} C.~M.,  et~al., 2021, \mn@doi [\mnras] {10.1093/mnras/stab1268},
  \href {https://ui.adsabs.harvard.edu/abs/2021MNRAS.504.5629R} {504, 5629}

\bibitem[\protect\citeauthoryear{{Raiteri} et~al.,}{{Raiteri}
  et~al.}{2023}]{raiteri2023}
{Raiteri} C.~M.,  et~al., 2023, \mn@doi [\mnras] {10.1093/mnras/stad942}, \href
  {https://ui.adsabs.harvard.edu/abs/2023MNRAS.522..102R} {522, 102}

\bibitem[\protect\citeauthoryear{{Rani}, {Valverde}  \& {La Mura}}{{Rani}
  et~al.}{2022}]{rani2022}
{Rani} B.,  {Valverde} J.,   {La Mura} G.,  2022, The Astronomer's Telegram,
  \href {https://ui.adsabs.harvard.edu/abs/2022ATel15375....1R} {15375, 1}

\bibitem[\protect\citeauthoryear{{Shukla} \& {Mannheim}}{{Shukla} \&
  {Mannheim}}{2020}]{shukla2020}
{Shukla} A.,  {Mannheim} K.,  2020, \mn@doi [Nature Communications]
  {10.1038/s41467-020-17912-z}, \href
  {https://ui.adsabs.harvard.edu/abs/2020NatCo..11.4176S} {11, 4176}

\bibitem[\protect\citeauthoryear{{Sironi}, {Petropoulou}  \&
  {Giannios}}{{Sironi} et~al.}{2015}]{sironi2015}
{Sironi} L.,  {Petropoulou} M.,   {Giannios} D.,  2015, \mn@doi [\mnras]
  {10.1093/mnras/stv641}, \href
  {https://ui.adsabs.harvard.edu/abs/2015MNRAS.450..183S} {450, 183}

\bibitem[\protect\citeauthoryear{{Stickel}, {Fried}, {Kuehr}, {Padovani}  \&
  {Urry}}{{Stickel} et~al.}{1991}]{stickel1991}
{Stickel} M.,  {Fried} J.~W.,  {Kuehr} H.,  {Padovani} P.,   {Urry} C.~M.,
  1991, \mn@doi [\apj] {10.1086/170133}, \href
  {http://adsabs.harvard.edu/cgi-bin/nph-bib_query?bibcode=1991ApJ...374..431S&db_key=AST}
  {374, 431}

\bibitem[\protect\citeauthoryear{{Stocke}, {Morris}, {Gioia}, {Maccacaro},
  {Schild}, {Wolter}, {Fleming}  \& {Henry}}{{Stocke}
  et~al.}{1991}]{stocke1991}
{Stocke} J.~T.,  {Morris} S.~L.,  {Gioia} I.~M.,  {Maccacaro} T.,  {Schild} R.,
   {Wolter} A.,  {Fleming} T.~A.,   {Henry} J.~P.,  1991, \mn@doi [\apjs]
  {10.1086/191582}, \href
  {https://ui.adsabs.harvard.edu/abs/1991ApJS...76..813S} {76, 813}

\bibitem[\protect\citeauthoryear{{Urry} \& {Padovani}}{{Urry} \&
  {Padovani}}{1995}]{urry1995}
{Urry} C.~M.,  {Padovani} P.,  1995, \mn@doi [\pasp] {10.1086/133630}, \href
  {https://ui.adsabs.harvard.edu/abs/1995PASP..107..803U} {107, 803}

\bibitem[\protect\citeauthoryear{{Valtaoja}, {L{\"a}hteenm{\"a}ki},
  {Ter{\"a}sranta}  \& {Lainela}}{{Valtaoja} et~al.}{1999}]{valtaoja1999}
{Valtaoja} E.,  {L{\"a}hteenm{\"a}ki} A.,  {Ter{\"a}sranta} H.,   {Lainela} M.,
   1999, \mn@doi [\apjs] {10.1086/313170}, \href
  {https://ui.adsabs.harvard.edu/abs/1999ApJS..120...95V} {120, 95}

\bibitem[\protect\citeauthoryear{{Villata} et~al.,}{{Villata}
  et~al.}{2002}]{villata2002}
{Villata} M.,  et~al., 2002, \mn@doi [\aap] {10.1051/0004-6361:20020662}, \href
  {https://ui.adsabs.harvard.edu/abs/2002A&A...390..407V} {390, 407}

\bibitem[\protect\citeauthoryear{{Villata} et~al.,}{{Villata}
  et~al.}{2006}]{villata2006}
{Villata} M.,  et~al., 2006, \mn@doi [\aap] {10.1051/0004-6361:20064817}, \href
  {https://ui.adsabs.harvard.edu/abs/2006A&A...453..817V} {453, 817}

\bibitem[\protect\citeauthoryear{{Villata} et~al.,}{{Villata}
  et~al.}{2008}]{villata2008}
{Villata} M.,  et~al., 2008, \mn@doi [\aap] {10.1051/0004-6361:200809552},
  \href {https://ui.adsabs.harvard.edu/abs/2008A&A...481L..79V} {481, L79}

\bibitem[\protect\citeauthoryear{{Villata} et~al.,}{{Villata}
  et~al.}{2009}]{villata2009}
{Villata} M.,  et~al., 2009, \mn@doi [\aap] {10.1051/0004-6361/200912732},
  \href {https://ui.adsabs.harvard.edu/abs/2009A&A...504L...9V} {504, L9}

\bibitem[\protect\citeauthoryear{{Vlasyuk}, {Spiridonova}, {Moskvitin}  \&
  {Maslennikova}}{{Vlasyuk} et~al.}{2022a}]{vlasyuk2022a}
{Vlasyuk} V.~V.,  {Spiridonova} O.~I.,  {Moskvitin} A.~S.,   {Maslennikova}
  O.~A.,  2022a, The Astronomer's Telegram, \href
  {https://ui.adsabs.harvard.edu/abs/2022ATel15344....1V} {15344, 1}

\bibitem[\protect\citeauthoryear{{Vlasyuk}, {Spiridonova}, {Moskvitin},
  {Emelianov}  \& {Fatkhullin}}{{Vlasyuk} et~al.}{2022b}]{vlasyuk2022b}
{Vlasyuk} V.~V.,  {Spiridonova} O.~I.,  {Moskvitin} A.~S.,  {Emelianov} E.~V.,
   {Fatkhullin} T.~A.,  2022b, The Astronomer's Telegram, \href
  {https://ui.adsabs.harvard.edu/abs/2022ATel15376....1V} {15376, 1}

\bibitem[\protect\citeauthoryear{{Wagner} \& {Witzel}}{{Wagner} \&
  {Witzel}}{1995}]{wagner1995}
{Wagner} S.~J.,  {Witzel} A.,  1995, \mn@doi [\araa]
  {10.1146/annurev.aa.33.090195.001115}, \href
  {https://ui.adsabs.harvard.edu/abs/1995ARA&A..33..163W} {33, 163}

\bibitem[\protect\citeauthoryear{{Wagner} et~al.,}{{Wagner}
  et~al.}{1993}]{wagner1993}
{Wagner} S.~J.,  et~al., 1993, \aap, \href
  {https://ui.adsabs.harvard.edu/abs/1993A&A...271..344W} {271, 344}

\bibitem[\protect\citeauthoryear{{Weaver} et~al.,}{{Weaver}
  et~al.}{2020}]{weaver2020}
{Weaver} Z.~R.,  et~al., 2020, \mn@doi [\apj] {10.3847/1538-4357/aba693}, \href
  {https://ui.adsabs.harvard.edu/abs/2020ApJ...900..137W} {900, 137}

\bibitem[\protect\citeauthoryear{{Zhang}, {Li}, {Guo}  \& {Giannios}}{{Zhang}
  et~al.}{2018}]{zhang2018}
{Zhang} H.,  {Li} X.,  {Guo} F.,   {Giannios} D.,  2018, \mn@doi [\apjl]
  {10.3847/2041-8213/aad54f}, \href
  {https://ui.adsabs.harvard.edu/abs/2018ApJ...862L..25Z} {862, L25}

\bibitem[\protect\citeauthoryear{{Zhang}, {Li}, {Giannios}, {Guo}, {Liu}  \&
  {Dong}}{{Zhang} et~al.}{2020}]{zhang2020}
{Zhang} H.,  {Li} X.,  {Giannios} D.,  {Guo} F.,  {Liu} Y.-H.,   {Dong} L.,
  2020, \mn@doi [\apj] {10.3847/1538-4357/abb1b0}, \href
  {https://ui.adsabs.harvard.edu/abs/2020ApJ...901..149Z} {901, 149}

\bibitem[\protect\citeauthoryear{{Zhang}, {Li}, {Giannios}, {Guo}, {Thiersen},
  {B{\"o}ttcher}, {Lewis}  \& {Venters}}{{Zhang} et~al.}{2022}]{zhang2022}
{Zhang} H.,  {Li} X.,  {Giannios} D.,  {Guo} F.,  {Thiersen} H.,
  {B{\"o}ttcher} M.,  {Lewis} T.,   {Venters} T.,  2022, \mn@doi [\apj]
  {10.3847/1538-4357/ac3669}, \href
  {https://ui.adsabs.harvard.edu/abs/2022ApJ...924...90Z} {924, 90}

\bibitem[\protect\citeauthoryear{{de Jaeger} et~al.,}{{de Jaeger}
  et~al.}{2023}]{dejaeger2023}
{de Jaeger} T.,  et~al., 2023, \mn@doi [\mnras] {10.1093/mnras/stad060}, \href
  {https://ui.adsabs.harvard.edu/abs/2023MNRAS.519.6349D} {519, 6349}

\makeatother
\end{thebibliography}

\section*{Affiliations}
{\it
$^{ 1}$INAF, Osservatorio Astrofisico di Torino, via Osservatorio 20, I-10025 Pino Torinese, Italy                                                                                                             \\
$^{ 2}$Saint Petersburg State University, 7/9 Universitetskaya nab., St. Petersburg, 199034 Russia                                                                                                             \\
$^{ 3}$Special Astrophysical Observatory of Russian Academy of Sciences, Nyzhnij Arkhyz, Karachai-Circassia, Russia, 369167                                                                                    \\
$^{ 4}$Pulkovo Observatory, St. Petersburg, 196140, Russia                                                                                                                                                     \\
$^{ 5}$Abastumani Observatory, Mt. Kanobili, 0301 Abastumani, Georgia                                                                                                                                          \\
$^{ 6}$University of Siena, Astronomical Observatory, Siena, Italy                                                                                                                                             \\
$^{ 7}$Astronomical Institute, Osaka Kyoiku University, Osaka 582-8582, Japan                                                                                                                                  \\
$^{ 8}$Hans-Haffner-Sternwarte (Hettstadt), Naturwissenschaftliches Labor f\"ur Sch\"uler, Friedrich-Koenig-Gymnasium, 97082 W\"urzburg, Germany                                                               \\
$^{ 9}$Department of Physics and Astronomy, N283 ESC, Brigham Young University, Provo, UT 84602, USA                                                                                                           \\
$^{10}$Section of Astrophysics, Astronomy and Mechanics, Department of Physics, National and Kapodistrian University of Athens, GR-15784 Zografos, Athens, Greece                                              \\
$^{11}$EPT Observatories, Tijarafe, La Palma, Spain                                                                                                                                                            \\
$^{12}$INAF, TNG Fundaci\'on Galileo Galilei, La Palma, Spain                                                                                                                                                  \\
$^{13}$Ulugh Beg Astronomical Institute, Astronomy Street 33, Tashkent 100052, Uzbekistan                                                                                                                      \\
$^{14}$Instituto de Astrof\'isica de Canarias (IAC), E-38200 La Laguna, Tenerife, Spain                                                                                                                        \\
$^{15}$Universidad de La Laguna (ULL), Departamento de Astrof\'isica, E-38206, La Laguna, Tenerife, Spain                                                                                                      \\
$^{16}$Instituto de Astrofísica de Andaluc\'ia (CSIC), Apartado 3004, E-18080 Granada, Spain                                                                                                                  \\
$^{17}$Institute of Astronomy and National Astronomical Observatory, Bulgarian Academy of Sciences, 72 Tsarigradsko shosse Blvd., 1784 Sofia, Bulgaria                                                         \\
$^{18}$Universidad Nacional Aut\'onoma de M\'exico, Instituto de Astronom\'ia, AP 70-264, CDMX 04510, Mexico                                                                                                   \\
$^{19}$Crimean Astrophysical Observatory RAS, P/O Nauchny, 298409, Russia                                                                                                                                      \\
$^{20}$Osservatorio Astronomico della Regione Autonoma Valle d’Aosta, I-11020 Nus, Italy                                                                                                                     \\
$^{21}$Institute of Astronomy, National Central University, Taoyuan 32001, Taiwan                                                                                                                              \\
$^{22}$Astronomical Observatory, Volgina 7, 11060 Belgrade, Serbia                                                                                                                                             \\
$^{23}$National University of Uzbekistan, Tashkent 100174, Uzbekistan                                                                                                                                          \\
$^{24}$Technische Universit\"at Dortmund, 44221 Dortmund, Germany                                                                                                                                              \\
$^{25}$INAF, Osservatorio Astrofisico di Catania, Via S.~Sofia 78, I-95123 Catania, Italy                                                                                                                      \\
$^{26}$Aryabhatta Research Institute of Observational Sciences (ARIES), Manora Peak, Nainital 263001, India                                                                                                    \\
$^{27}$Universidad Nacional Aut\'onoma de M\'exico, Instituto de Astronom\'ia, AP 106, Ensenada 22800, Baja California, Mexico                                                                                 \\
$^{28}$Department of Physics and Astronomy, Faculty of Natural Sciences, University of Shumen, 115, Universitetska Str., 9712 Shumen, Bulgaria                                                                 \\
$^{29}$Department of Physics and Electronics, CHRIST (Deemed to be University), Hosur Main Road, Bengaluru - 560029, India                                                                                     \\
$^{30}$Institute of Astrophysics, Foundation for Research and Technology - Hellas, Vasilika Vouton, GR-70013 Heraklion, Greece                                                                                 \\
$^{31}$Department of Physics, University of Crete, Voutes University Campus, GR-70013 Heraklion, Greece                                                                                                        \\
$^{32}$Engelhardt Astronomical Observatory, Kazan Federal University, Tatarstan, Russia                                                                                                                        \\
$^{33}$Universit\"at W\"urzburg, 97074 W\"urzburg, Germany                                                                                                                                                     \\
$^{34}$Dip. di Scienze Fisiche, della Terra e dell’Ambiente, Universit\`a di Siena, via Roma 56, 53100, Siena, Italy                                                                                         \\
$^{35}$Department of Physics, Astronomy and Geophysics, Connecticut College, New London, CT 06320, USA                                                                                                         \\
$^{36}$AAVSO observer, Russia                                                                                                                                                                                  \\
$^{37}$Remote observer of Burke-Gaffney Observatory, Canada                                                                                                                                                    \\
$^{38}$Owens Valley Radio Observatory, California Institute of Technology, MC 249-17, Pasadena, CA 91125, USA                                                                                                  \\
 }

\bsp	
\label{lastpage}
\end{document}